\documentclass[journal]{IEEEtran}
\usepackage{cite}
\usepackage{amsmath,amssymb,amsfonts}
\usepackage{algorithmic}
\usepackage{graphicx}
\usepackage{textcomp}
\usepackage{xcolor}
\usepackage{subcaption}
\usepackage{multirow}
\usepackage{ifpdf}
\DeclareUnicodeCharacter{223C}{-}
\DeclareUnicodeCharacter{2212}{-}
\DeclareUnicodeCharacter{03B2}{-}

\ifCLASSINFOpdf
  \usepackage{graphicx}
\fi
\usepackage{array}
\begin{document}

%\input{JSSC.bbl}
%\bibliographystyle{IEEEtran}
%\bibliography{reference}
%\bstctlcite{IEEEexample:BSTcontrol}
\title{Multi-Mode Spatial Signal Processor with Rainbow-like Fast Beam Training and Wideband Communications using True-Time-Delay Arrays}
%\title{A 61dB SNDR Sub-0.4V Pulse Biased VCO-based Readout Circuit with Time-Domain Offset Calibration Enabled by Natural Energy Harvesting}

\author{Chung-Ching~Lin,~\IEEEmembership{Graduate Student Member,~IEEE,} 
        Chase~Puglisi,~\IEEEmembership{Member,~IEEE,} 
        Veljko~Boljanovic,~\IEEEmembership{Graduate Student~Member,~IEEE,}
        Han~Yan,~\IEEEmembership{Graduate~Student Member,IEEE,}        
        Erfan~Ghaderi,~\IEEEmembership{Member,IEEE,} 
        Jayce~Gaddis,~\IEEEmembership{Student Member,IEEE}
        Qiuyan~Xu,~\IEEEmembership{Graduate Student Member,IEEE,}
        Sreeni~Poolakkal,~\IEEEmembership{Graduate Student Member,~IEEE,}
        Danijela~Cabric,~\IEEEmembership{Fellow, IEEE,}
        and~Subhanshu~Gupta,~\IEEEmembership{Senior Member,~IEEE}

\thanks{This work was supported in part by NSF under grants 1955672, 1705026, and 1944688. This work was also supported in part by the ComSenTer and CONIX Research Centers, two of six centers in JUMP, a Semiconductor Research Corporation (SRC) program sponsored by DARPA.}%
\thanks{Chung-Ching Lin, Erfan Ghaderi, Chase Puglisi, Jayce Gaddis, and Subhanshu Gupta are with the School of Electrical Engineering and Computer Science, Washington State University, Pullman, WA 99164 USA (e-mail: chung-ching.lin@wsu.edu).}% <-this % stops a space
\thanks{Veljko Boljanovic, Han Yan, and Danijela Cabric are with the Department of Electrical and Computer Engineering, University of California, Los Angeles, Los Angeles, CA 90095 USA (e-mail:
vboljanovic@ucla.edu).}% <-this % stops a space

%\thanks{Manuscript received July 2020;.}
}

\maketitle
\vspace{-5mm}

\begin{abstract}
Initial access in millimeter-wave (mmW) wireless is critical toward successful realization of the fifth-generation (5G) wireless networks and beyond. Limited bandwidth in existing standards and use of phase-shifters in analog/hybrid phased-antenna arrays (PAA) are not suited for these emerging standards demanding low-latency direction finding. This work proposes a reconfigurable true-time-delay (TTD) based spatial signal processor (SSP) with frequency-division beam training methodology and wideband beam-squint less data communications. Discrete-time delay compensated clocking technique is used to support 800~MHz bandwidth with a large unity-gain bandwidth ring-amplifier (RAMP)-based signal combiner. To extensively characterize the proposed SSP across different SSP modes and frequency-angle pairs, an automated testbed is developed using computer-vision techniques that significantly speeds up the testing progress and minimize possible human errors. Using seven levels of time-interleaving for each of the 4 antenna elements, the TTD SSP has a delay range of 3.8~ns over 800~MHz and achieves unique frequency-to-angle mapping in the beamtraining mode with nearly 12 dB frequency-independent gain in the beamforming mode.  The SSP is prototyped in 65nm CMOS with an area of 1.98mm$^2$ consuming only 29~mW excluding buffers. Further, an error vector magnitude (EVM) of 9.8\% is realized for 16-QAM modulation at a speed of 122.8~Mb/s.
\end{abstract}

\begin{IEEEkeywords}
Fast beam training, frequency-to-angle mapping, true-time-delay spatial signal processing, ring amplifier.
\end{IEEEkeywords}

% \fi
\IEEEpeerreviewmaketitle

\section{Introduction}
\IEEEPARstart{A}{bundant} spectral resources in the millimeter wave regime have opened new possibilities to deliver Gb/s data. Owing to the large propagation losses at mmW, multi-antenna systems are needed for improved signal-to-noise ratio (SNR) and link quality. However, the resulting narrow-pointed beams necessitate fast beam acquisition especially for communications applications on-the-move.  To acquire fast channel state information (CSI), maximize the received power and fully benefit from the beamforming technology, a process known as beamtraining is widely adopted to search the optimal angles to direct the beams. 

\begin{figure}[t]
\centering
    \includegraphics[width=1\columnwidth]{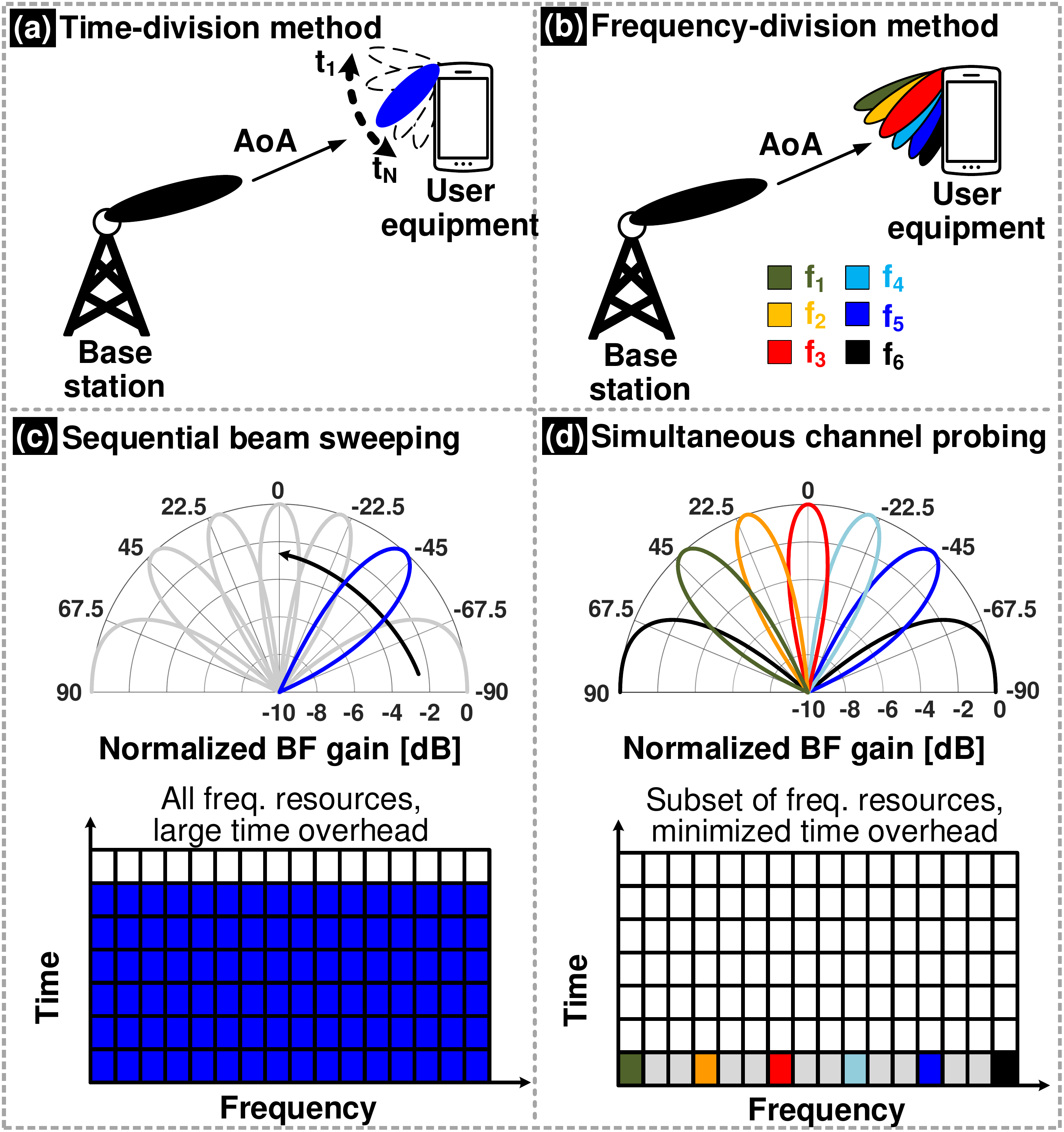}
    \vspace{-4mm}
    \caption{\small (a) Time-division beam sweep, (b) frequency-division beam sweep; and their anticipated beam patterns and resource allocation over both frequency and time shown in (c) and (d).}
    \vspace{-7mm}
    \label{fig:fig1}
\end{figure}

\begin{figure}[t]
\centering
    \includegraphics[width=0.9\columnwidth]{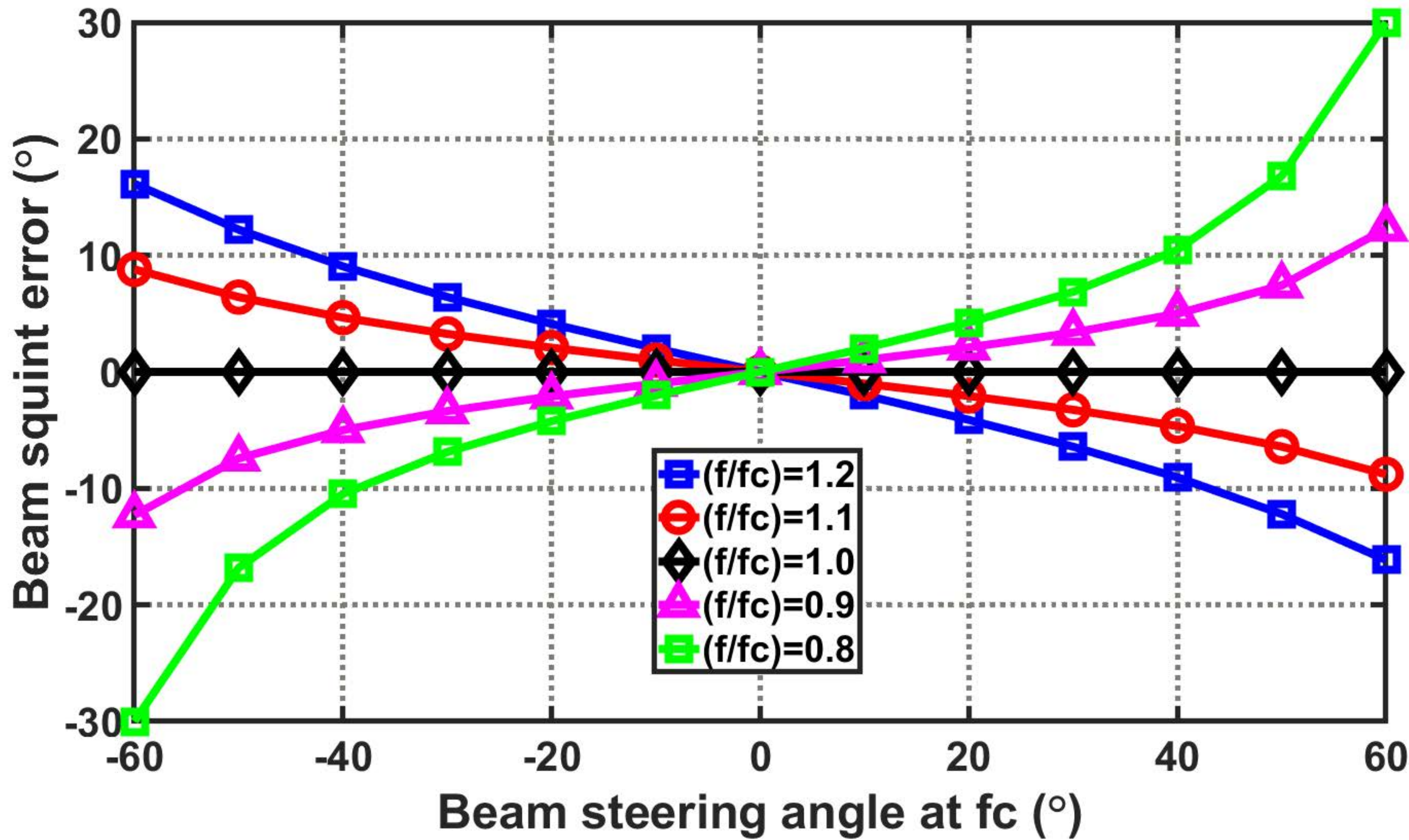}
    \vspace{-2mm}
    \caption{\small Beam-squint effect  using phase-shifters based PAAs.}
    \vspace{-3mm}
    \label{fig:fig2}
\end{figure}

Existing beam training methodologies are based on iterative scanning approach limited to non-scalable, time- and/or power-consuming scenarios and pose numerous challenges toward low-latency requirement for both the base station (BS) and the user equipment (UE). Fig.~\ref{fig:fig1}(a) illustrates the time-division based beam scanning method used in most of analog phased antenna arrays (PAA). By configuring the phase shifters inherent to these arrays and sensing the received power, the CSI can be obtained. Finer angle resolution is limited by aperture that increase the design complexity and consumes large power and silicon area \cite{pepe_2017}. In contrast to the time-division based beam scanning method, Fig.~\ref{fig:fig1}(b) shows the frequency-division based beam scanning approach that obtains the CSI by observing different array frequency responses within a single orthogonal frequency-division multiplexing (OFDM) symbol  saving critical time needed for beam scanning. Nevertheless, the need for a constant group delay in the frequency-division based beam training limits its realization when using  phase shifters \cite{yan_2019}. The phase shift approximation of the inter-element time delay holds true over a small fractional bandwidth only leading to large errors at the band edges and degraded signal-to-noise ratio  as the channel bandwidth scales up as shown in Fig.~\ref{fig:fig2}. 

Uniform-frequency response across wide modulated bandwidths can be realized using the concept of true-time-delay (TTD) introduced with baseband delay elements. Large delay range-to-resolution ratios realized using this technique \cite{ghaderi_tcas1_2019,ghaderi_jssc_2021} has opened opportunities for fast beam training algorithms in large-scale arrays without needing additional external constructs. For example, Rotman Lens \cite{gao_2017} based beamforming can be leveraged for beamtraining. However, its angular resolution is being limited to three points and its integration adds costs and makes the system bulky. Benefiting from several GHz of available spectrum, another recent approach used two Leaky Wave Antennas (LWA) \cite{saeidi_2021} operating at sub-THz frequencies was used for directional finding. LWA at lower frequency band has also been proposed in \cite{li_2021} but with a coverage efficiency of only 50\%.  Like the previous approach, integration of these antennas can be costly and inefficient especially when operating in the cm- and mm-wave bands.  

The authors' recent works have shown a reconfigurable TTD-based SSP with large delay-bandwidth product that creates unique frequency-to-angle map for fast beamtraining and reduces the scan time of analog PAAs from several symbols to a single OFDM symbol \cite{yan_2019,boljanovic_2020,boljanovic_2021}. In \cite{lin_2021}, the authors demonstrated a reconfigurable TTD SSP hardware for the first  time that can be easily reconfigured for both beamtraining and beamforming  over 800~MHz modulated bandwidth. This work significantly expands on the circuits and systems proposed in \cite{lin_2021} with the following distinct contributions: 
\begin{itemize}
    \item	In-depth analysis of the hardware architecture when applying the beamtraining algorithm, 
    \item	Hardware design considerations and requirements for beam training algorithm are presented and compared with various TTD implementations from RF to analog/digital basebands, 
    \item	Detailed circuit design of the TTD SSP including time-interleaved interpolated clock generator and high-speed sampling technique to alleviate jitter and mismatch, and 
    \item	Expanded measurement methodology detailing the test setup with customized automated validation bench development for single-tone, wideband, and modulated measurements of the TTD SSP.
\end{itemize}

The rest of this article is organized as follows. Section II discusses system design considerations for the two SSP modes outlining hardware challenges. The proposed SSP design and circuit implementation details are presented in Section III. Section IV presents the measurement results for different SSP modes followed by conclusions in Section V.

\section{TTD SSP System Design Considerations}
\label{sec:sec2}
We consider a uniform linear array with N critically spaced antenna elements. The inter-element delay can be expressed as $\Delta t = sin(\theta)/2/f_c$, where $\theta$ is the angle of incidence, and f\textsubscript{c} is the center frequency. Therefore, the array response for the n$^{th}$ element in frequency domain is $[a(\theta,f)]_n=exp(-j \cdot 2\pi(n-1)d\cdot f \cdot sin(\theta)/c)$, \textcolor{black}{where $d$ is the inter-element spacing, $f$ is the frequency, and $c$ represents the speed of light\cite{lin_2021}.} Next, we describe the system design considerations for both wideband beamtraining and data communications mode with associated design parameters and hardware challenges. 

\subsection{Mode 1: Low-latency beamtraining}
\begin{figure}[t]
\centering
    \includegraphics[width=1\columnwidth]{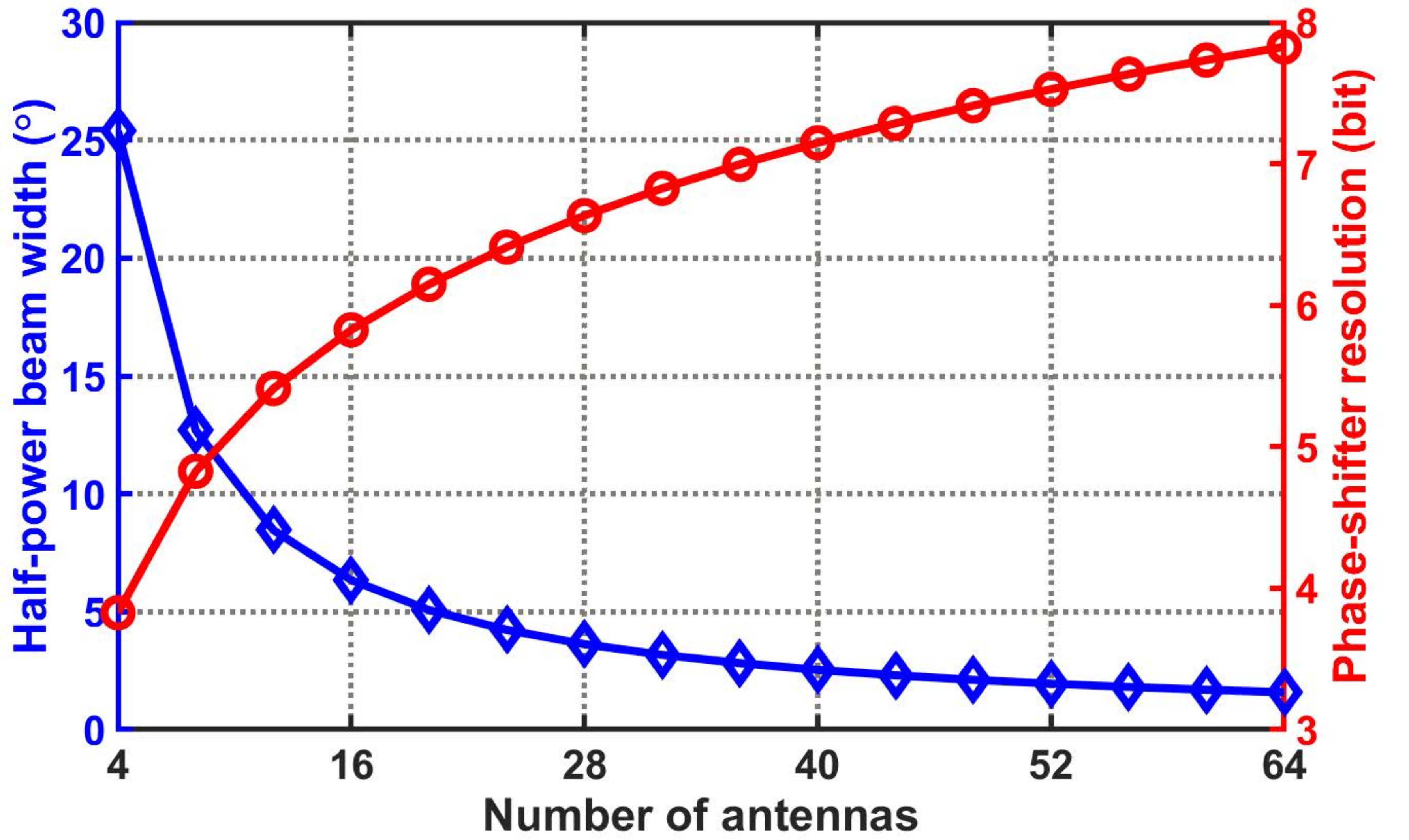}
    \vspace{-4mm}
    \caption{\small Half-power beam width and estimated phase-shifter resolution against array size for beam-training in phase-shifter based PAAs. }
    \vspace{-6mm}
    \label{fig:fig3}
\end{figure}

Most of the existing beamtraining algorithms leverage only phase-shifter based SSPs for exhaustive beam sweeping where different beams are used sequentially in time to estimate the angle-of-arrival (AoA). These algorithms (and consequently, the analog PAAs) suffer from two main obstacles and are thus limited to smaller array sizes by the number of antennas and the phase-shifters complexity. Fig.~\ref{fig:fig3} quantifies the required phase-shifter resolution (i.e., complexity) \textcolor{black}{to cover an entire 360$^{\circ}$ plane} using half-power beam width (HPBW) with respect to the array size assuming a linear array architecture.  As the array size increases, the HPBW drops which implies the resolution of the phase-shifters also needs to be increased to cover the entire 360$^\circ$range. Considering similar bandwidth, loss, and root-mean-square gain/phase error, increasing the phase-shifter resolution is often accompanied with high power consumption and silicon area \cite{pepe_2017} adding prohibitive hardware overhead in the PAAs. 

To alleviate the above problems, the authors introduce architectures that can synthesize rainbow-like beams to detect the entire angular range in a single OFDM symbol \cite{yan_2019,boljanovic_2021}. The key idea is to place different signal delays in different antenna branches to exacerbate beam squint intentionally among subcarriers. By properly configuring the delay taps in an array, each frequency can be associated with a unique beam that creates a unique frequency-to-angle map. The CSI and the AoA can be observed by its spectral analysis embedded in the signal spectrum obtained simply through frequency-domain DSP.  

\begin{table}
\footnotesize {
\caption{Time-delay unit implementation choices.}
\label{tab:tab1}
\setlength{\tabcolsep}{4pt}
\centering
\begin{tabular}{|c|c|c|c|c|} 
\hline
\multirow{2}{*}{}                  & \textbf{Area}    & \textbf{Power}    & \textbf{Delay}    & \textbf{Delay}\\ 
%\cline{2-9}
                                   &  & \textbf{Cons.} & \textbf{Range} & \textbf{Resolution}  \\ 
\hline
\hline
\textbf{xLine}    &   High      &   Low ($\approx$0)  &   Low ($\approx$0)   &   Limited \\ 
\hline
\textbf{LC delay line}    &  High      &   Low ($\approx$0)  &   Low ($\approx$0)   &   Limited \\
\hline
\textbf{RF resampling} & Low &   Med. &   Med. &   High      \\ 
\hline
\textbf{Gm-C filter}                & Med.              & Med.-to-High               & Med.            & Med.             \\ 
\hline
\textbf{BB sampling}  &   Low & Low-to-med. & High & High\\ 
\hline
\textbf{Digital}                  & Low-to-med.  & High    & High  & High  \\
\hline
\hline

\end{tabular}
} \\
\end{table}

In this work, when TTD delay are set to be $\tau_\textsubscript{n}$ in the n$^{th}$ element of a uniform linear array, different incident angle $\theta$ results in different system frequency response that can be represented as $|W^H(f)a(\theta,f)|^2$, where the $W(f)$ is the antenna weight factor,  $\theta$ is the angle-of-incidence, and \textit{f} is the frequency. Hence, one can infer unknown $\theta$ by measuring the received frequency response formed from a multi-carrier pilot symbol. Furthermore, with the signal bandwidth, BW, it has been proved in \cite{boljanovic_2021} that setting the TTD delay taps as $\tau_\textsubscript{n}=(n-1)/BW, n=1,\dots,N$ enables searching the entire angular range $\theta \in [−\pi/2, \pi/2]$ at once. The approach, however, requires the maximum TTD delay to be $\tau_\textsubscript{N} = (N-1)/BW$ at the last antenna and hence poses a significant design challenge for array implementation. It is also worth to mention briefly that the algorithm can be further enhanced by increasing its frequency diversity order, R. The delay taps are $\tau_\textsubscript{n} = R \cdot (n - 1)/BW, n=1,\dots,N$, which maps R different subcarriers in a particular probed direction. In our implementation, a unity R is used for proof-of-concept. 

\subsection{Mode 2: Wideband communications}
\begin{figure}[t]
\centering
    \includegraphics[width=1\columnwidth]{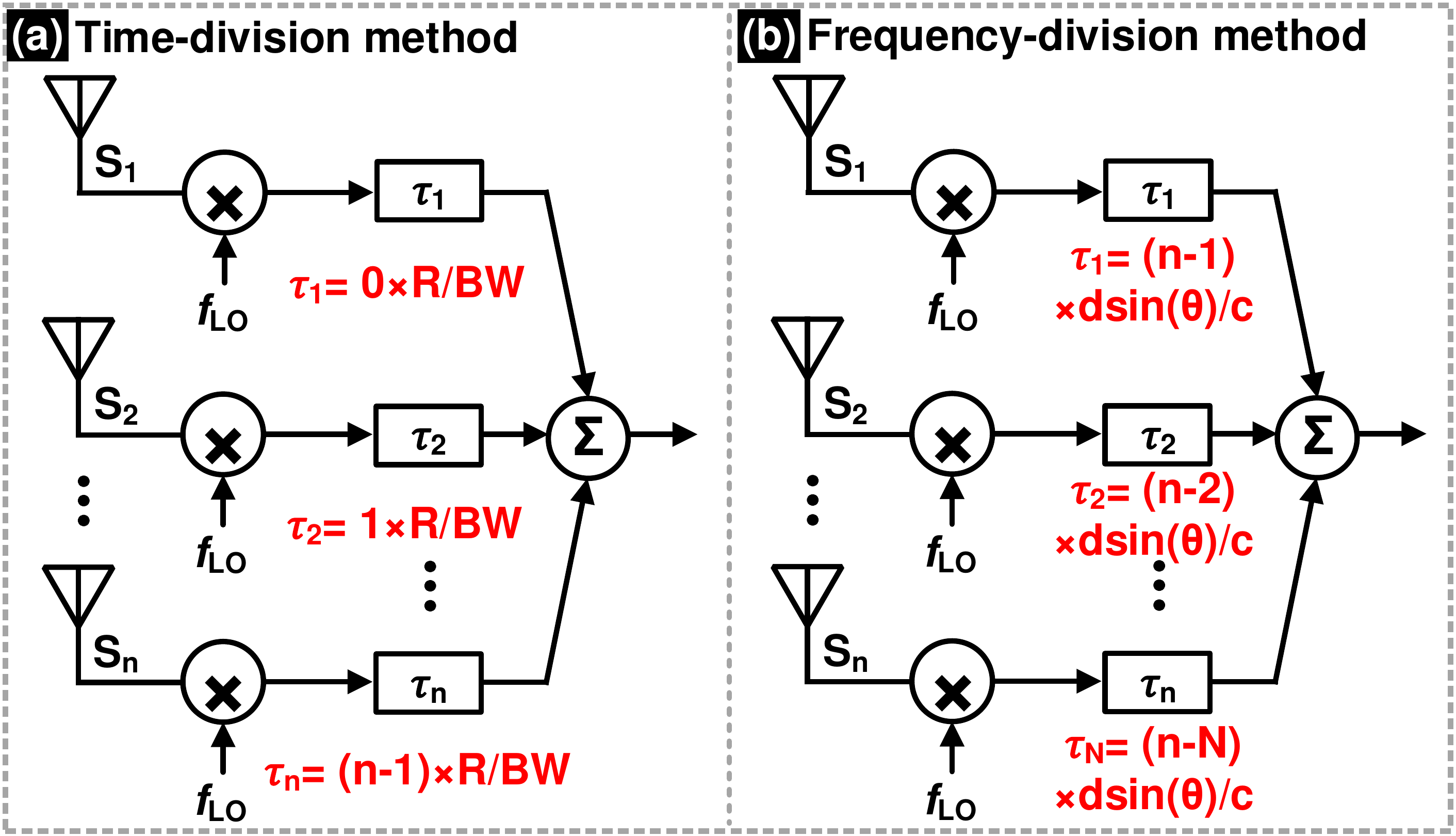}
    \vspace{-4mm}
    \caption{\small Delay tap requirement of (a) beamtraining; and (b) beamforming mode. }
    \vspace{-6mm}
    \label{fig:fig4}
\end{figure}

After identifying the AoA using the beamtraining algorithm described in the previous subsection, the SSP can be switched from beamtraining mode to beamforming mode for data communication. The received signals at the antennas are aligned first by applying suitable time delays and phase-shifts (in baseband) and then constructively combined enhancing the SNR by N over a wideband frequency. Different from the beamtraining mode, the delay taps $\tau_n$ need to be designed such that antenna weight, $W(f)$, matches with array spatial response, $a(\theta, f)$, for all frequencies which requires the delay taps $\tau_n = (n-1)\cdot d\cdot sin(\theta)/c, n = 1,\cdots N$. Figure~\ref{fig:fig4} illustrates the tap delay setting for both the modes that are used to characterize our proposed multi-mode TTD SSP.

\begin{figure}[t]
\centering
    \includegraphics[width=1\columnwidth]{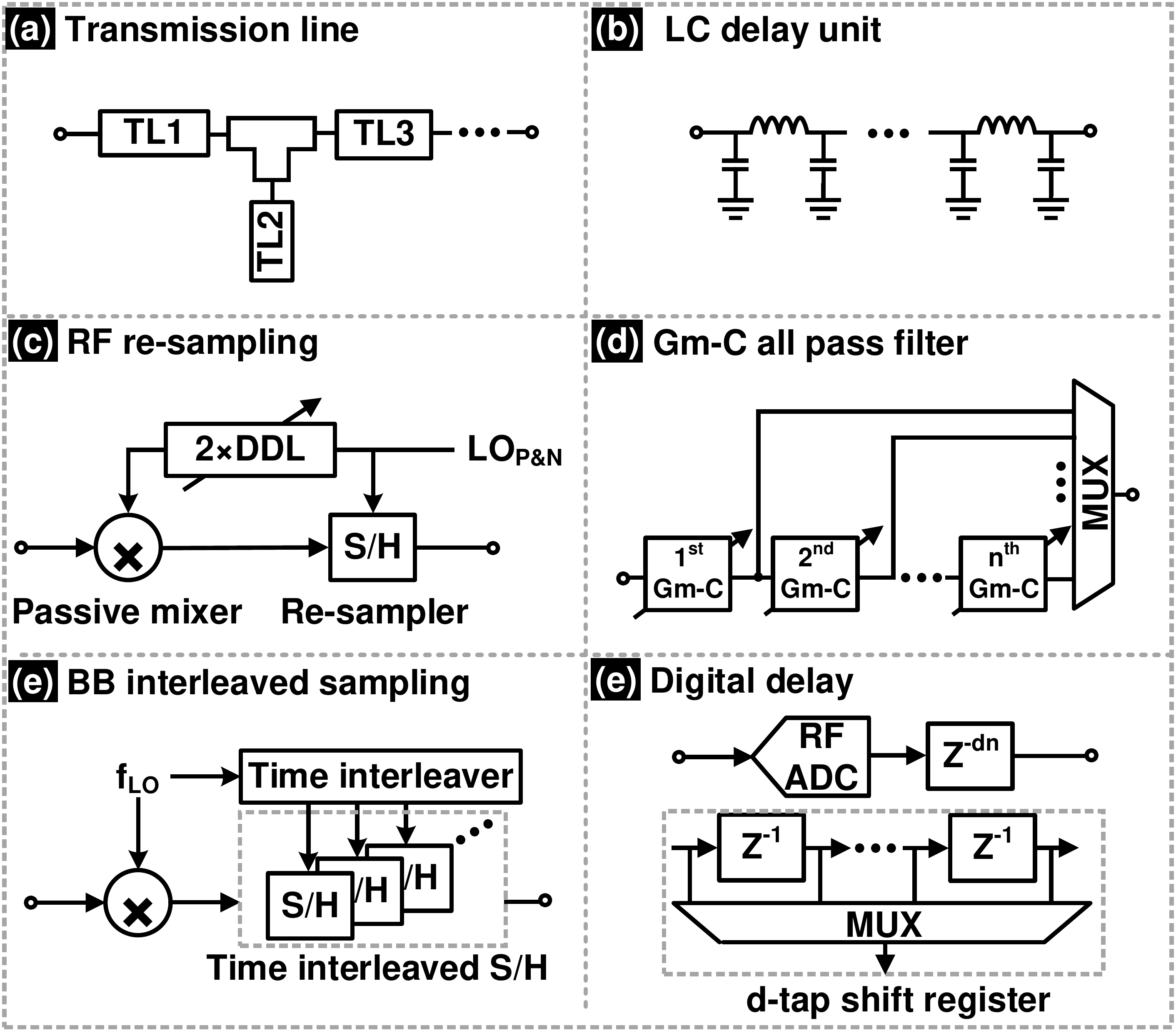}
    \vspace{-4mm}
    \caption{\small Methods to implement time delay units (TDU): (a) microwave transmission line, (b) LC delay line, (c) RF resampling, (d) Gm-C filter, (e) baseband (BB) sampling, and (f) digital domain. }
    \vspace{-6mm}
    \label{fig:fig5}
\end{figure}

%%%%******************************* *********************************************%%%%
%%%%%            SECTION III
%%%%%%%%%%%%%%%%%%%%%%%%%%%%%%%%%%%%%%%%%%%%%%%%%%%%%%%%%%%%%%%%%%%%%%%%%%%%%%%%%%%%%%%%%%%%%%%%%
\section{Proposed TTD SSP Circuit Implementation}
\label{sec:sec3}
As discussed in Section~\ref{sec:sec2}, the integration of the rainbow-like beamtraining algorithm with the data communications mode requires a reconfigurable TTD architecture with both constant group delay and a large delay range-to-resolution ratio across a wide bandwidth to enable the two SSP modes.

\begin{figure*}[t]
\centering
    \includegraphics[width=1\textwidth]{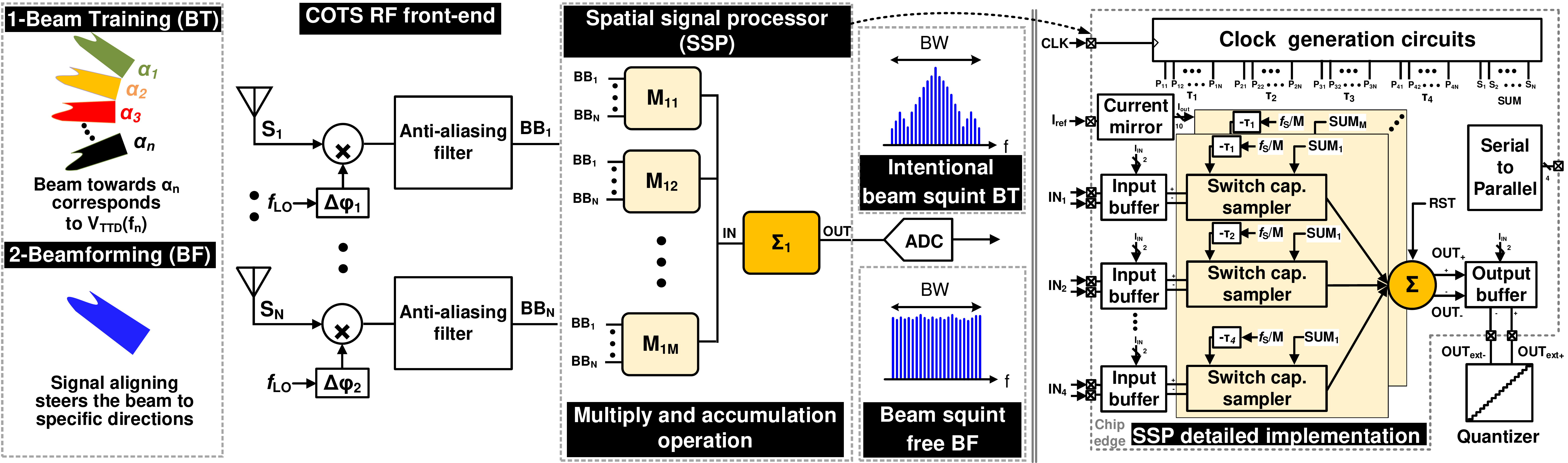}
    \vspace{-4mm}
    \caption{Proposed SSP modes; architecture; and block-level implementation (inset). }
    \vspace{-6mm}
    \label{fig:fig6}
\end{figure*}

Recent time delay units (TDU) have used either of the following methods: transmission line \cite{lin_2019}, LC delay network \cite{hashemi_2008,li_2020}, RF re-sampling \cite{spoof_2020}, Gm-C all pass filter \cite{garakoui_2015,mondal_2017}, time interleaved switched capacitor circuit \cite{ghaderi_tcas1_2019,ghaderi_jssc_2021, nagulu_2021}, and digital delays \cite{jang_2019}. The conceptual implementations of these TDUs are shown in Fig.~\ref{fig:fig5}(a)-(f). Table~\ref{tab:tab1} further provides a high-level comparison of these methods. As seen in Table~\ref{tab:tab1}, 
the transmission line and LC delay network provide a passive solution however with limited delay resolution and delay range. The delay range can be improved by switching sections of the TDUs and cascading multiple sections but requires a large silicon area. Moderate resolution and delay range can be realized using Gm-C delay cell. Nevertheless, as the bandwidth increases, the power consumption of the single Gm-C unit also increases proportionally. Besides, the Gm-C filter hardly benefit from process scaling due to the analog-intensive architecture. The process, voltage, and temperature effects can be mitigated by including additional calibration knobs for biasing input pairs or load at the expense of additional power and bandwidth constraint owing to the larger parasitic capacitance.  The RF re-sampling \cite{spoof_2020} is performed by using a digital-controlled delay lines (DDL) to resample the input signal after a passive mixer which demonstrate a wide delay range with fine resolution. However, interleaving directly at carrier frequency is potentially prone to jitter and gain errors. Besides, the achievable resolution is determined by the DDL design that also operates at carrier frequency implying the design challenges and power overhead also scales up with frequency. Another alternative is to introduce delay in the digital domain using the powerful DSP. However, the need for an ADC per antenna element results in significant power overhead increasing proportionately for larger arrays. Furthermore, integer delay can be implemented in digital domain easily while fractional delay is achieved by interpolation which incurs additional power/design overhead \cite{jang_2019}. In contrast, baseband sampling \cite{ghaderi_tcas1_2019,ghaderi_tmtt_2020,nagulu_2021} using interleaved sampler has a high delay range-to-resolution ratio while the higher delay range is achieved by increasing the interleaving factor. The inherent nature of digital-mostly architecture of the time-interleaved sampling technique with precision on-chip capacitor make this architecture friendly to process scaling.

Fig.~\ref{fig:fig6} shows the proposed the TTD SSP with detailed architecture implementation. The front-end of the system is expected to operate with a center frequency (f\textsubscript{c}) of 28~GHz and a target bandwidth (BW) of 800~MHz. The received signal is then down converted to baseband, assuming a low-IF downconverter.  Two functionalities including beamtraining and beamforming are demonstrated through the frequency-independent time delay settings with a multiply-and-accumulated (MAC) operation. The results are observed directly from the output power spectral density. The MAC structure can also be scaled to construct a higher order matrix in a MIMO system with minimal hardware overhead. 
\textcolor{black}{Leveraging the the baseband discrete-time beamforming, it has been proved in \cite{ghaderi_tcas1_2019} that the minimum number of interleaving levels (\textit{M}) to meet the required signal bandwidth (\textit{BW}) can be expressed as (assuming a 120$^\circ$ field-of-view (FoV)):}
\begin{equation}
    \vspace{-1mm}
    \label{eq:eq1}
    M=1+[(d/\lambda_\textsubscript{c})\times \sqrt{3}/2 \times (N-1) \times (BW/f_\textsubscript{c})]
    \vspace{-1mm}
\end{equation}
where %\textit{M} is the minimum number of interleaving factor, BW represents the signal bandwidth,
\textit{d} is the inter-element distance, $\lambda$\textsubscript{c} is the wavelength,  and f\textsubscript{c} is the center frequency. \textcolor{black}{Using (\ref{eq:eq1})  and considering critically-spaced antennas with a BW of 800~MHz with f\textsubscript{c} half of its BW, an interleaving level of 7 is required to support a 4-element SSP implementation.} 

The next sub-sections presents the implementation of key components.

\begin{figure}[t]
\centering
    \includegraphics[width=1\columnwidth]{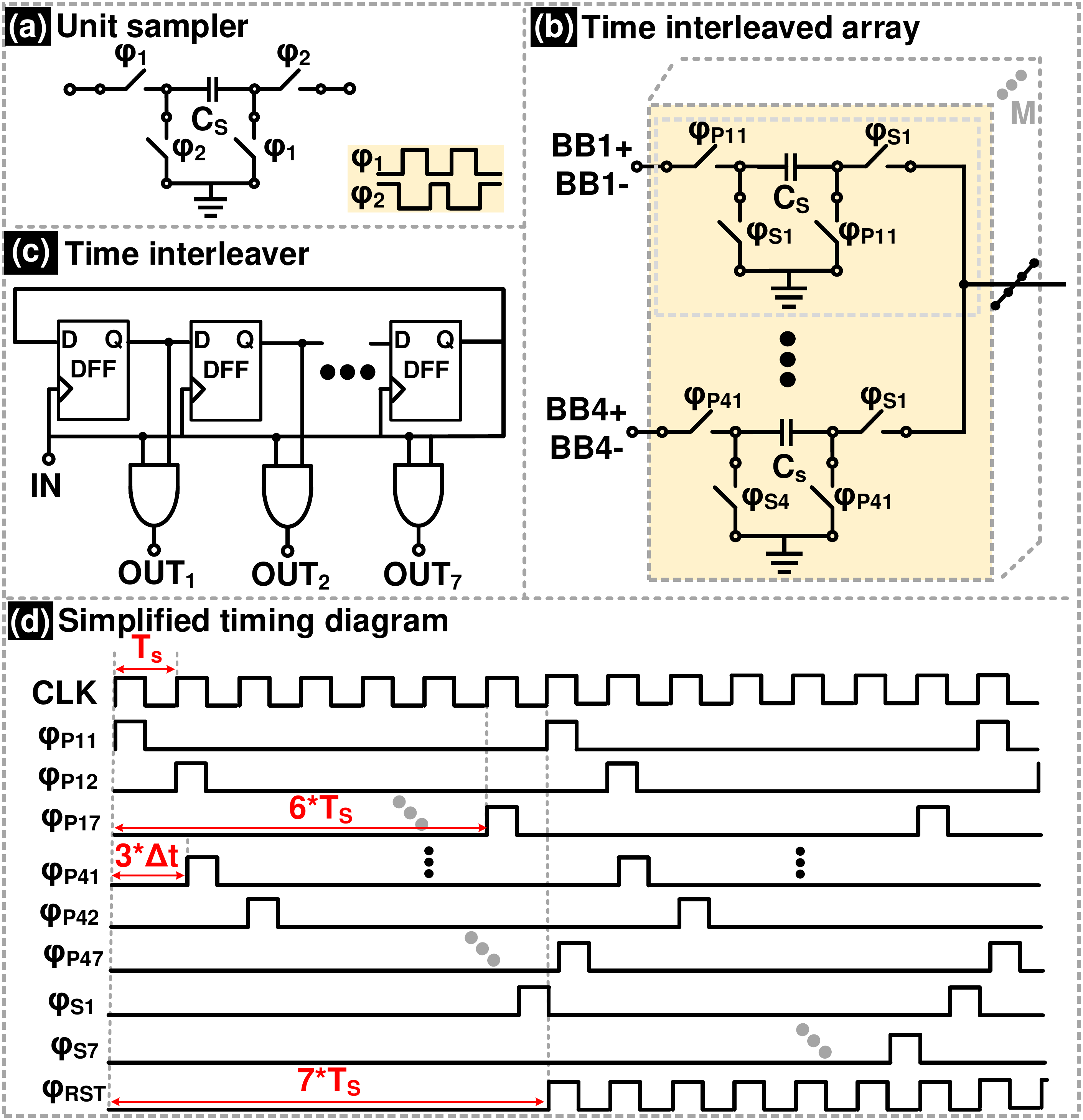}
    \vspace{-4mm}
    \caption{\small (a) Unit sampler, (b) time-interleaved (TI) sampler array, (c) time interleaver, and (d) simplified timing   diagram.}
    \vspace{-6mm}
    \label{fig:fig7}
\end{figure}

\subsection{Switched capacitor array (SCA) and clock generation}
The fundamental block of the discrete-time delay compensation method is the sample-and-hold within the MAC. The sampler, shown in Fig.~\ref{fig:fig7}(a), is adopted from \cite{allstot_1978} as it offers greater immunity to parasitic capacitance thus alleviating second-order errors. The complete sampler array is illustrated in Fig.~\ref{fig:fig7}(b) constructed using the unit sampler in Fig.~\ref{fig:fig7}(a). The SCA requires non-overlapping clocks for sampling, sum, and reset phases generated by the time-interleaver. Each interleaved level has a conversion speed of roughly 228~MHz (=f\textsubscript{clk}/M). After all the signal paths are aligned in time, the sampled inputs from the 4-elements are combined using a closed-loop ring amplifier (RAMP) before being fed into the quantizer (off-chip) for further processing. The parasitic-insensitive topology is highly beneficial as the interleaved sampled signals from different elements are summed at the opamp virtual ground. The choice of the sampling capacitor, Cs, (=50fF) is determined by the thermal noise requirement and the available silicon area.  The required time interleaved clock is derived from the input clock (IN) and derived by the circuits illustrated in Fig.~\ref{fig:fig7}(c). The timing diagram shown in Fig.~\ref{fig:fig7}(d) captures the relationship between sampling, summing, and reset phases with twice the number of interleaving levels and eight times the bandwidth. Careful layout with mismatch optimized post-layout extractions was done to alleviate mismatch in the time-interleaved clock phases similar to a high-speed time-interleaved ADC \cite{hershberg_2012}.

\begin{figure}[t]
\centering
    \includegraphics[width=1\columnwidth]{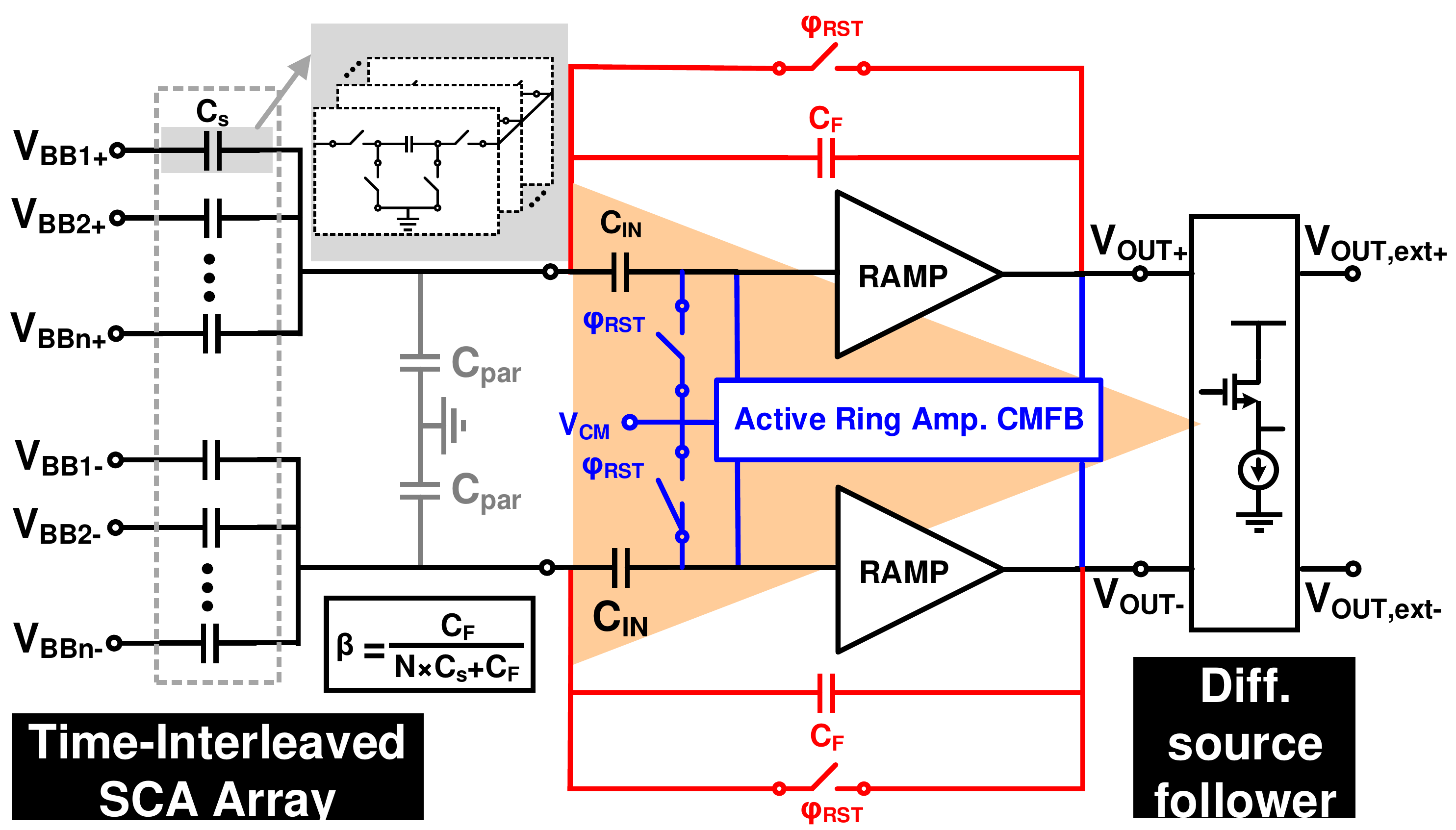}
    \vspace{-4mm}
    \caption{\small (a) Unit sampler, (b) time-interleaved (TI) sampler array, (c) time interleaver, and (d) simplified timing   diagram.}
    \vspace{-4mm}
    \label{fig:fig8}
\end{figure}

\begin{figure}[t]
\centering
    \includegraphics[width=1\columnwidth]{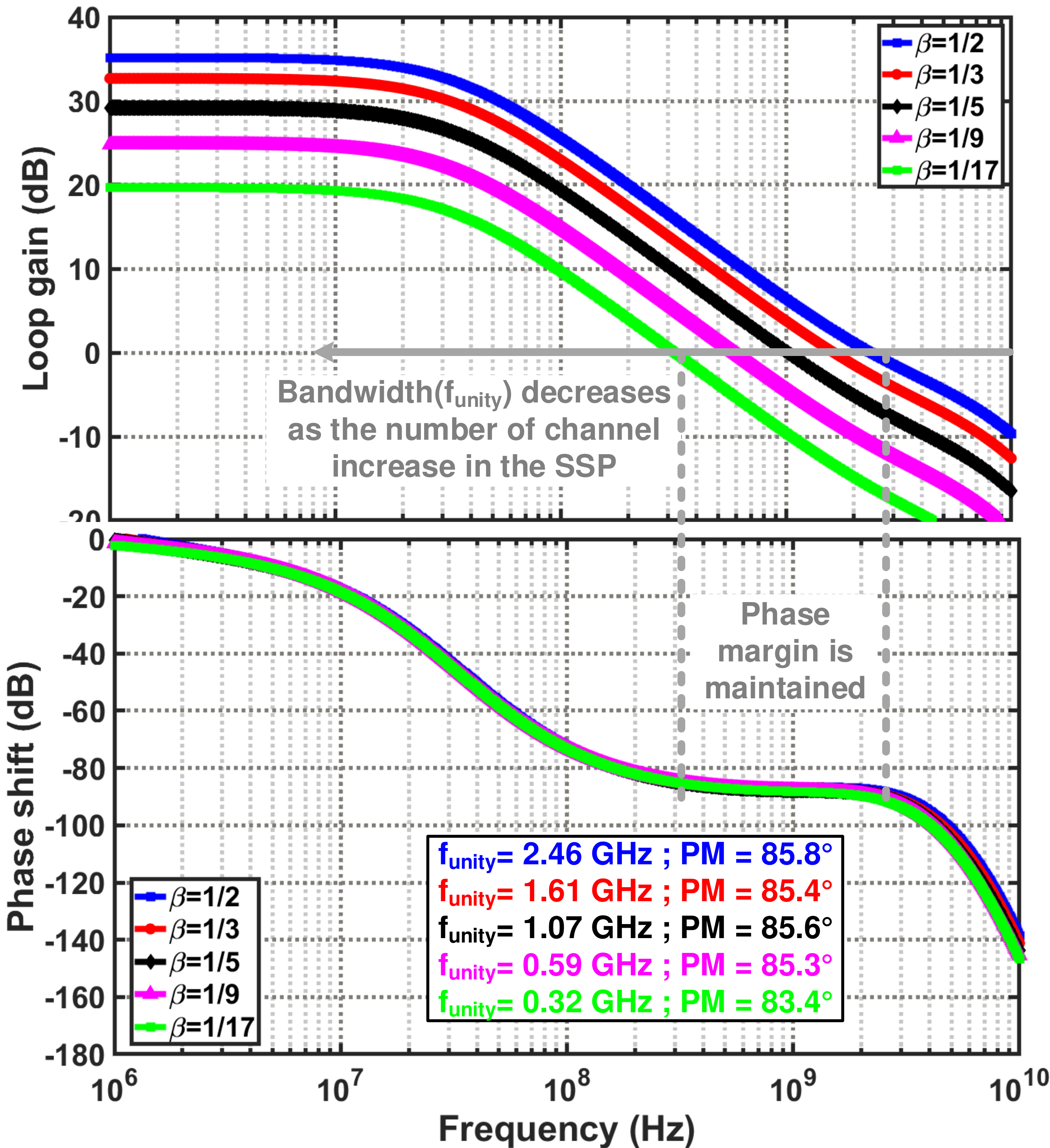}
    \vspace{-4mm}
    \caption{\small Simulated loop gain and phase response for different feedback factor ($\beta$).}
    \vspace{-8mm}
    \label{fig:fig9}
\end{figure}

\subsection{Signal combiner}
The large feedback factors due to summation of different antenna inputs and interleaved channels at the OTA virtual ground significantly constrains the unity-gain bandwidth affecting settling times for wideband signals and therefore resulting in significant power/performance penalties. Emerging research in RAMP topology \cite{hershberg_2012,lim_2015,lagos_2019,lim_jssc_2015,lagos_1_2019} alleviates some of these limitations enabling realization of a low-voltage high-speed amplifier that scales well with advanced CMOS technology. A RAMP comprises of several cascaded inverters that form a multi-stage amplifier. Its initial response behaves analogous to a ring oscillator to charge and discharge the capacitive load which is stabilized by placing a dominant output pole, through adjusting the biasing points and hence creating a dead-zone at the last stage of the RAMP.  The dead-zone enables the RAMP to operate the last stage in either subthreshold or cutoff which increases its output impedance pushing the dominant pole (at the output) towards the origin thus stabilizing the amplifier.  

Fig.~\ref{fig:fig8}(a) shows the top-level open-loop dynamic amplifier design adapted from \cite{lagos_2019}. It consists of two single-ended RAMP that forms a pseudo-differential configuration and eventually construct a closed loop integrator with additional feedback capacitor and a reset switch. The bias enhancement and the biasing scheme using an anti-parallel (AP) arrangement of CMOS transistors are implemented here to enhance both the bandwidth and the linearity. Though the RAMP behavior has been investigated thoroughly in \cite{hershberg_2012,lagos_2019,lagos_1_2019}, the application of the RAMP in our SSP has additional challenges. Recall from Fig.~\ref{fig:fig6} and Fig.~\ref{fig:fig8}, during the SUM phase, the SSP is analogous to a closed loop integrator. As the number of elements increase, the feedback factor, $\beta$, decreases which implies the available loop gain is inversely proportional to the number of elements. In practice, additional parasitic capacitance located at the virtual ground node of the RAMP due to routing the time-interleaved channels makes $\beta$ even smaller. Fig.~\ref{fig:fig9} shows the simulated loop gain and phase for the RAMP with different β maintained for all the cases. Thus, the adoption of the RAMP is crucial here compared to the traditional OTA to achieve large delay range across a wide bandwidth. Besides, the first- and second-stage of the RAMP are made as fast as possible, and eventually process limited. An advanced technology node will further help to achieve better gain-bandwidth products alleviating the $\beta$ constraints. Lastly, similar to \cite{lagos_1_2019}, the active common-mode feedback (CMFB) using 2-stage RAMP is used to define the common-mode voltage providing sufficient gain compared to a passive CMFB.

 \begin{figure}[t]
\centering
    \includegraphics[width=1\columnwidth]{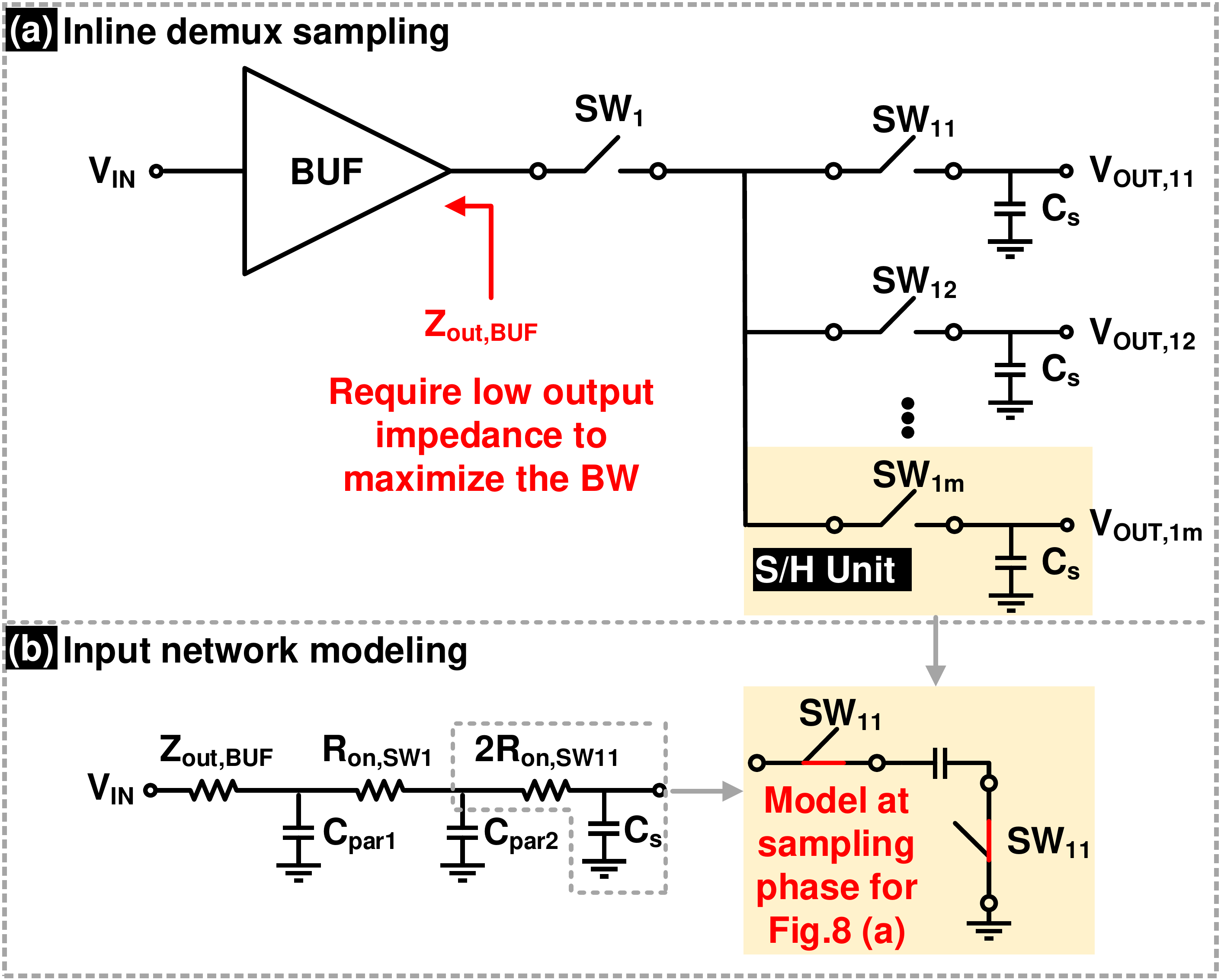}
    \vspace{-4mm}
    \caption{\small (a) Inline demux sampling used in the proposed SSP; and (b) simplified RC network for BW prediction.}
    \vspace{-6mm}
    \label{fig:fig10}
\end{figure}

 \begin{figure}[t]
\centering
    \includegraphics[width=1\columnwidth]{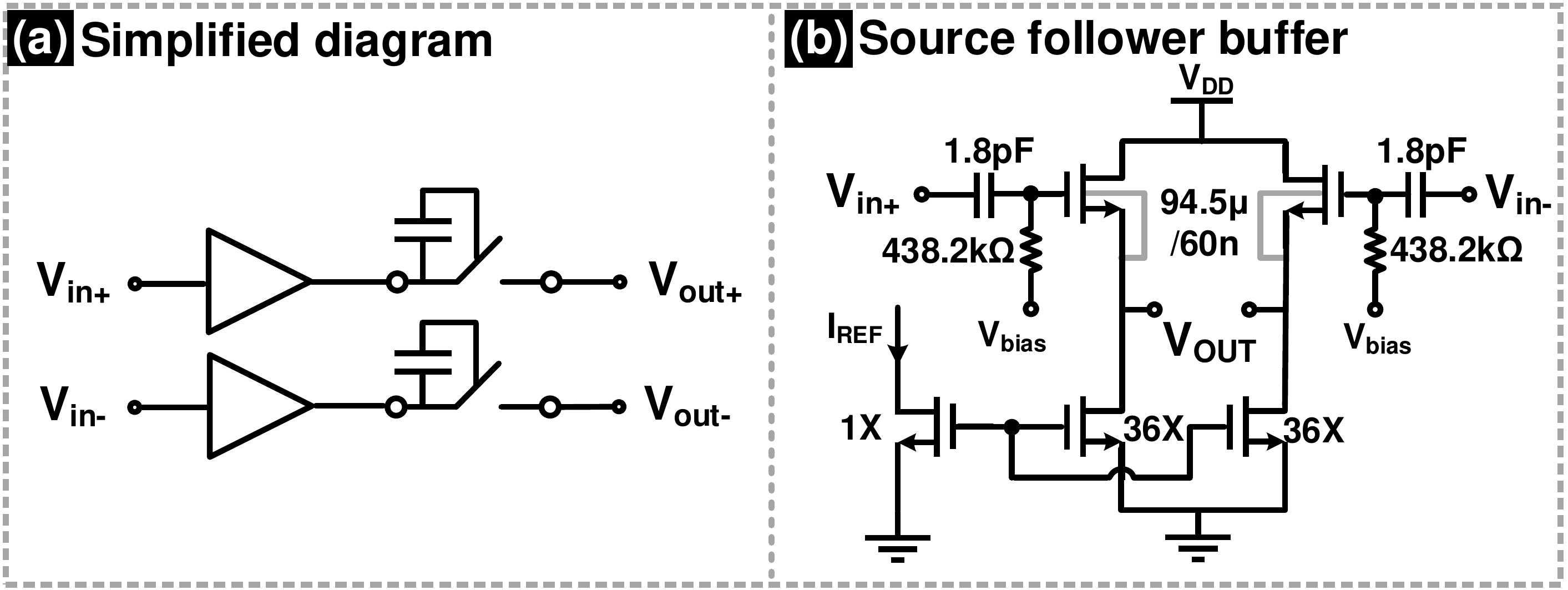}
    \vspace{-4mm}
    \caption{\small Implementation of the input network (a) simplified diagram; (b) differential source follower buffer.}
    \vspace{-6mm}
    \label{fig:fig11}
\end{figure}
 
 \begin{figure}[t]
\centering
    \includegraphics[width=1\columnwidth]{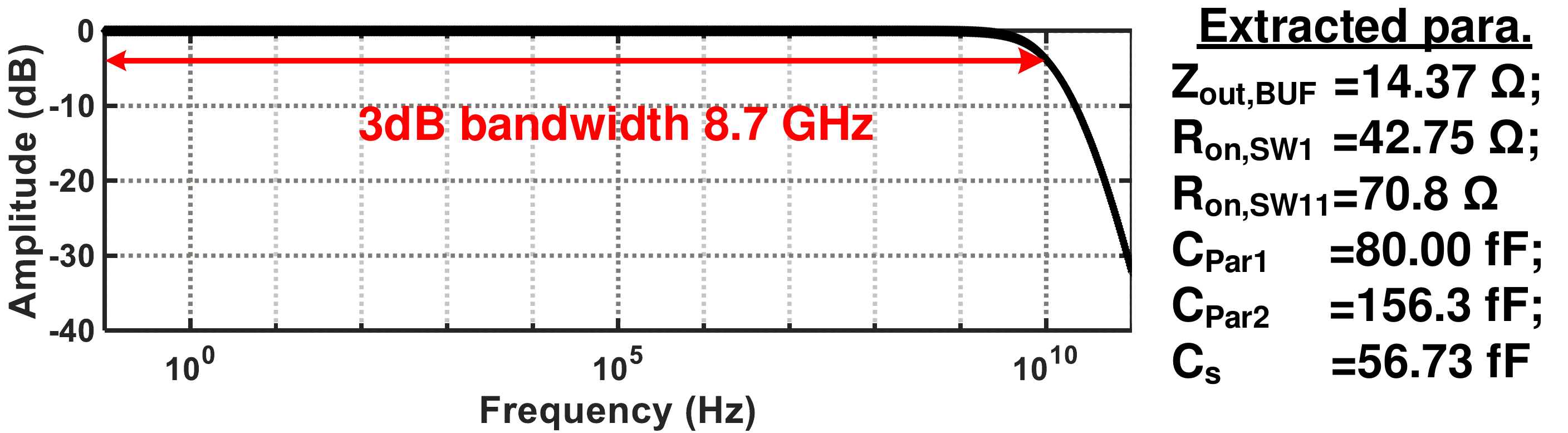}
    \vspace{-4mm}
    \caption{\small Amplitude response of the input network using extracted parameters.}
    \vspace{-6mm}
    \label{fig:fig12}
\end{figure}

 \begin{figure}[t]
\centering
    \includegraphics[width=1\columnwidth]{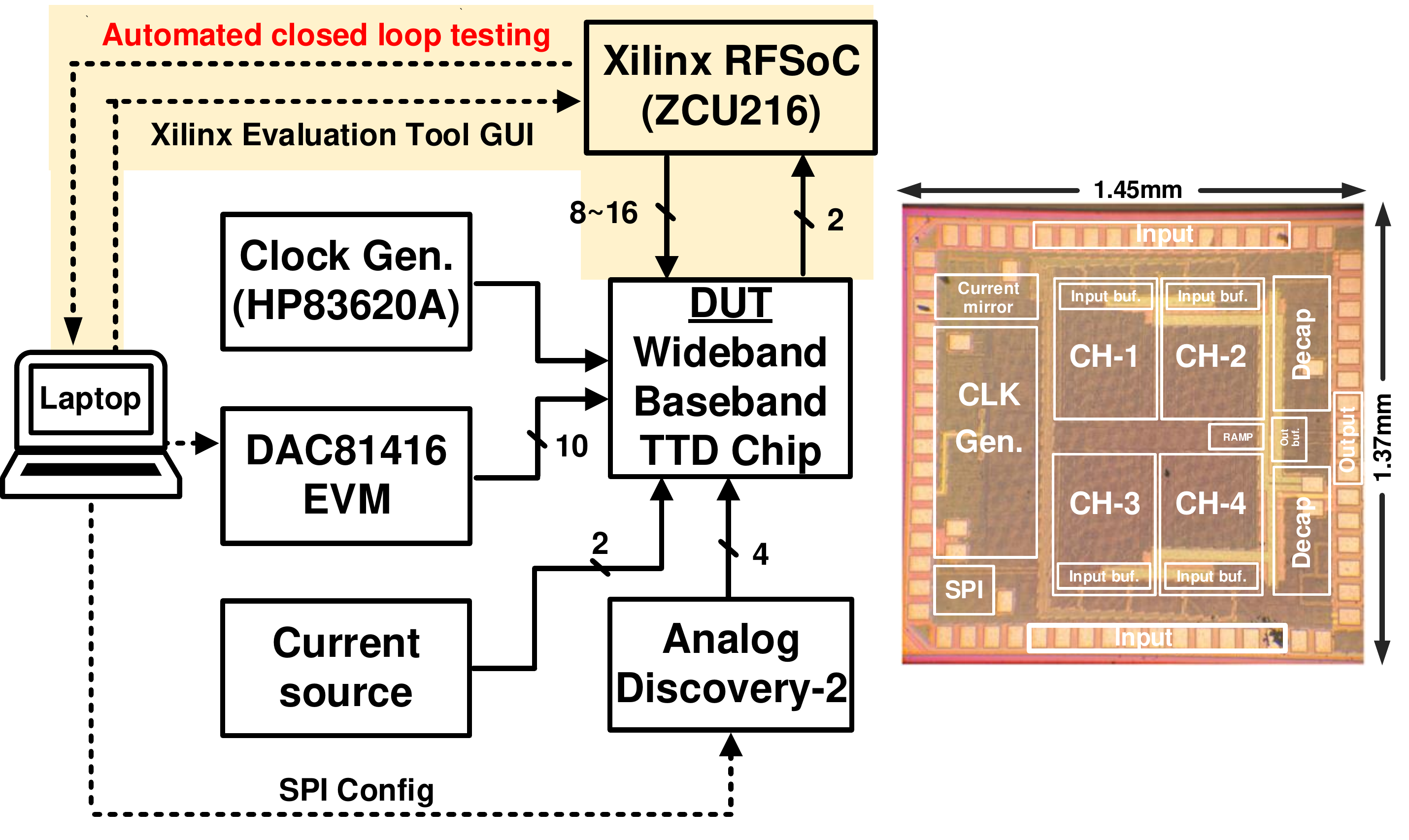}
    \vspace{-4mm}
    \caption{\small TTD SSP automated closed-loop for characterization and chip micrograph.}
    \vspace{-4mm}
    \label{fig:fig13}
\end{figure}

\subsection{Input and output buffers}
The input network including the sample-and-hold also determine the bandwidth and linearity of the SSP in addition to the signal combiner. Fig.~\ref{fig:fig10}(a) shows the time-interleaved sampling scheme used in the proposed SSP input network. Its RC model is illustrated in Fig.~\ref{fig:fig10}(b) to predict the bandwidth where Z\textsubscript{out,BUF} is the buffer output impedance, R\textsubscript{on} represents the switch on-resistance, and C\textsubscript{par} is the node parasitic capacitance including the switches’ capacitance.  To achieve fast settling for wideband signals, input buffers with low output impedance are required. Thus, a source follower is designed as the input buffer to provide a low output impedance. Following the buffer, the sampling switch (SW\textsubscript{1}) needs to handle the entire bandwidth-of-interest with high linearity and is thus designed as a boosted switch. 

\begin{figure*}[t]
\centering
    \includegraphics[width=0.8\textwidth]{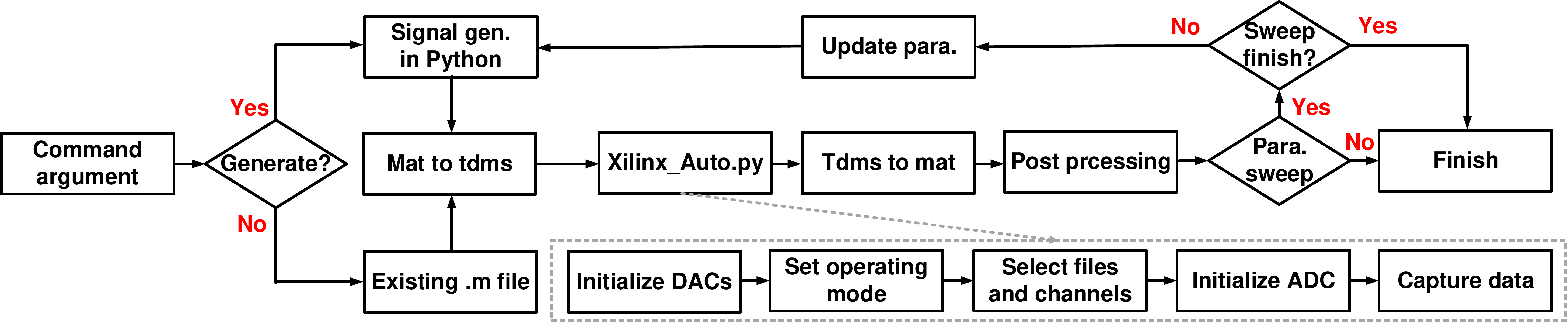}
    \vspace{-2mm}
    \caption{Automated testing flow for multi-mode TTD SSP characterization.  }
    \vspace{-6mm}
    \label{fig:fig14}
\end{figure*}

The last section of the RC network is determined by two principles. First, the interleaved switches are triggered using the non-overlapping clocks with duty cycle of 7.14\% (1/14), given an interleaving factor of 7 where only one switch is off at a particular interval and decide the network bandwidth.  Second, the size of the sampling capacitor C\textsubscript{s} is constrained by thermal noise requirements with the noise floor. Thus, the bandwidth will be determined by two times the on-resistance of SW\textsubscript{11}. Based on the above design considerations, C\textsubscript{s} is chosen to be 50~fF, while the interleaved switches are implemented using transmission gates considering the input level requirement and its R\textsubscript{on} is almost independent of the output voltages. The simplified diagram and transistor-level 
schematic of the critical blocks is shown in Fig.~\ref{fig:fig11}(a)-(b) respectively. Fig.~\ref{fig:fig11}(b) shows the differential buffers starting with a bias tee network for biasing the input buffer input transistor independent to the external signal. The bootstrap switch is adapted from \cite{abo_1999} for the SSP input network providing a low on-resistance of $42.75\Omega$. Other parameters in Fig.~\ref{fig:fig11}(b) are extracted from simulation (not considering routing parasitic) demonstrating a flat amplitude response with 3dB bandwidth of 8.7~GHz meeting settling requirements over bandwidth-of-interest (800~MHz), as shown in Fig.~\ref{fig:fig12}. Note, the high pass network formed by the bias tee is omitted in the bandwidth estimation. The output buffer design is similar to the input buffer except the Gm of the input pair is set as 20~mS to drive the $50\Omega$ interface for data acquisition.

\section{Testbed Automation and Measurement Results}
\label{sec:sec4}
The proposed TTD SSP is designed and fabricated in a TSMC 65 nm CMOS occupying an area of 1.98 mm\textsuperscript{2} including test pads as shown in Fig.~\ref{fig:fig13} and mounted on a prototyped  evaluation board for testing. Fig.~\ref{fig:fig13} also shows the testbed used to characterize the TTD SSP.   

Multi-antenna SSP receivers are typically characterized over a large matrix of frequency and angle-of-arrival combinations requiring significant effort complicated by multiple inputs/outputs. Thus, the development of the test bench and validation method is of interest. The input signal for the TTD SSP device-under-test (DUT) was generated in MATLAB and then uploaded to a Xilinx ZCU216 RFSoC configured in multi-tier synchronization mode to precisely synchronize the 8-differential channels (16 inputs) before being applied to the DUT through SMA coaxial cables. A 3.2~GHz clock source from HP8360A provides the external clock to generate the required time-interleaved phases on-chip. The required biases and configuration control are provided by the Texas Instruments (TI) DAC (DAC81416EVM) and Digilent Analog Discovery. Finally, the output signal is fed to the GS/s ADCs in the Xilinx RFSoC and observed from its GUI. 

Because the input and output signal are generated and acquired by the same GUI, signal path automation is possible to reduce the testing time. For precise synchronization with multi-channel generation, we leverage computer vision techniques to test GUI with pre-stored image icons. The testbench specifies which GUI components to interact with. PyAutoGui library is used to provide an abstraction to OpenCV \cite{chang_2010} which provides tools to easily distinguish the screen images. Figure~\ref{fig:fig14} shows the flow chart for the automated closed loop testbed. The input can be selected from either pre-generated MATLAB code or directly in Python. Another application script called as Xilinx\_Auto.py configures the DACs and read the captured ADC data for additional post processing.

\subsection{Mode 1: Low-latency beamtraining test}
\begin{figure}[t]
\centering
    \includegraphics[width=1\columnwidth]{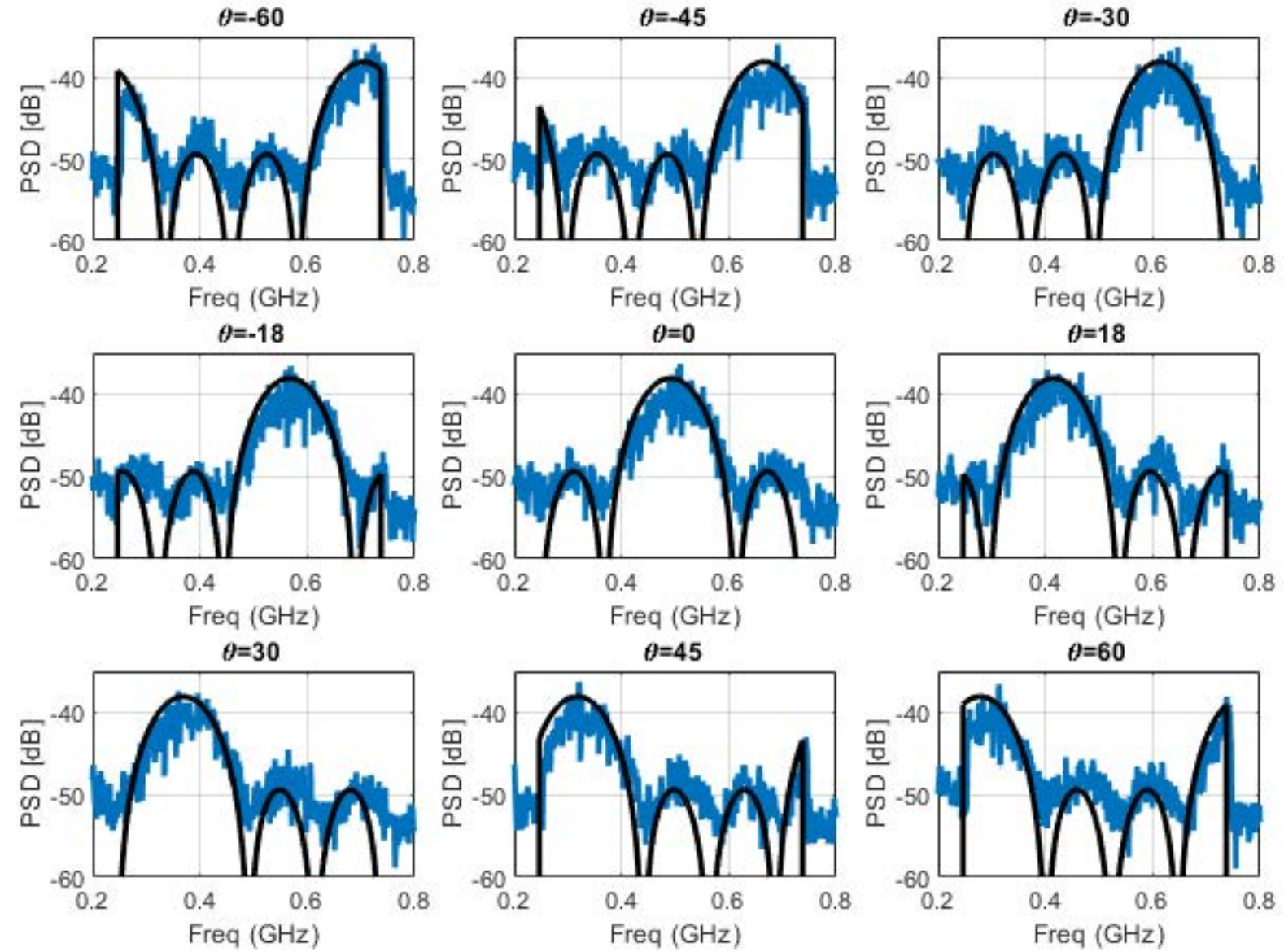}
    \vspace{-4mm}
    \caption{\small Measured beamtraining power spectral density (PSD) against theoretical results. }
    \vspace{-6mm}
    \label{fig:fig15}
\end{figure}

\begin{figure}[t]
\centering
    \includegraphics[width=1\columnwidth]{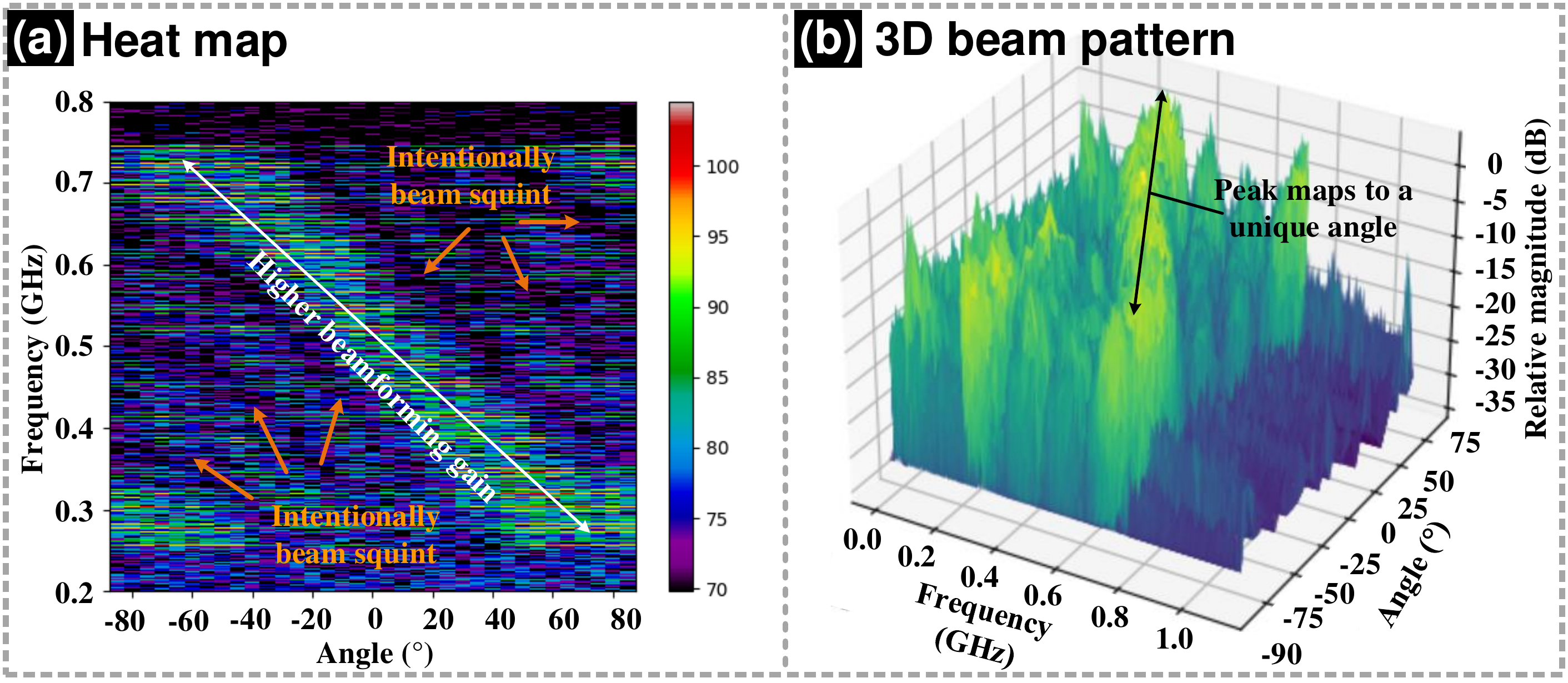}
    \vspace{-4mm}
    \caption{\small Measured (a) heat map; and (b) 3D beam pattern for beamtraining.}
    \vspace{-5mm}
    \label{fig:fig16}
\end{figure}

In the beamtraining mode, the input signals of the TTD SSP are generated by RF DAC which emulates OFDM symbols received by a critically spaced linear array at $\lambda/2$ with incident angle $\theta$ followed by down-conversion to intermediate frequency with a BW of 491.32~MHz and subcarrier spacing 960~kHz. Although the beamtraining conceptually requires only one pilot OFDM symbol to measure the system frequency response and infer $\theta$, our experiment lacks the necessary synchronization in the current test bench setup. Thus, the same OFDM pilot symbol repeats itself and its power spectral density (PSD) is observed for analysis. It is clear from the measured response in Fig.~\ref{fig:fig15} that the PSD of the received signal is uniquely determined by incident angles $\theta$. Hence, $\theta$ can be inferred without time consuming sequential beam sweep. To further prove the effectiveness of the beamtraining algorithm, Fig.~\ref{fig:fig16}(a) shows the measured heat map with a sweep from -85$^\circ$ to 85$^\circ$, constructing a unique frequency-to-angle mapping within the high beamforming gain region. The dark color region represents the intentionally squint region which matches the single-point measurement shown in Fig. ~\ref{fig:fig15}. Another way to interpret the measured data is leveraging the three-dimensional plot among frequency, angle, and relative magnitude, illustrated in Fig.~\ref{fig:fig16}(b). Again, a peak (i.e., high beamforming gains region) is clearly observed across different angles. 

 \begin{figure}[t]
\centering
    \includegraphics[width=1\columnwidth]{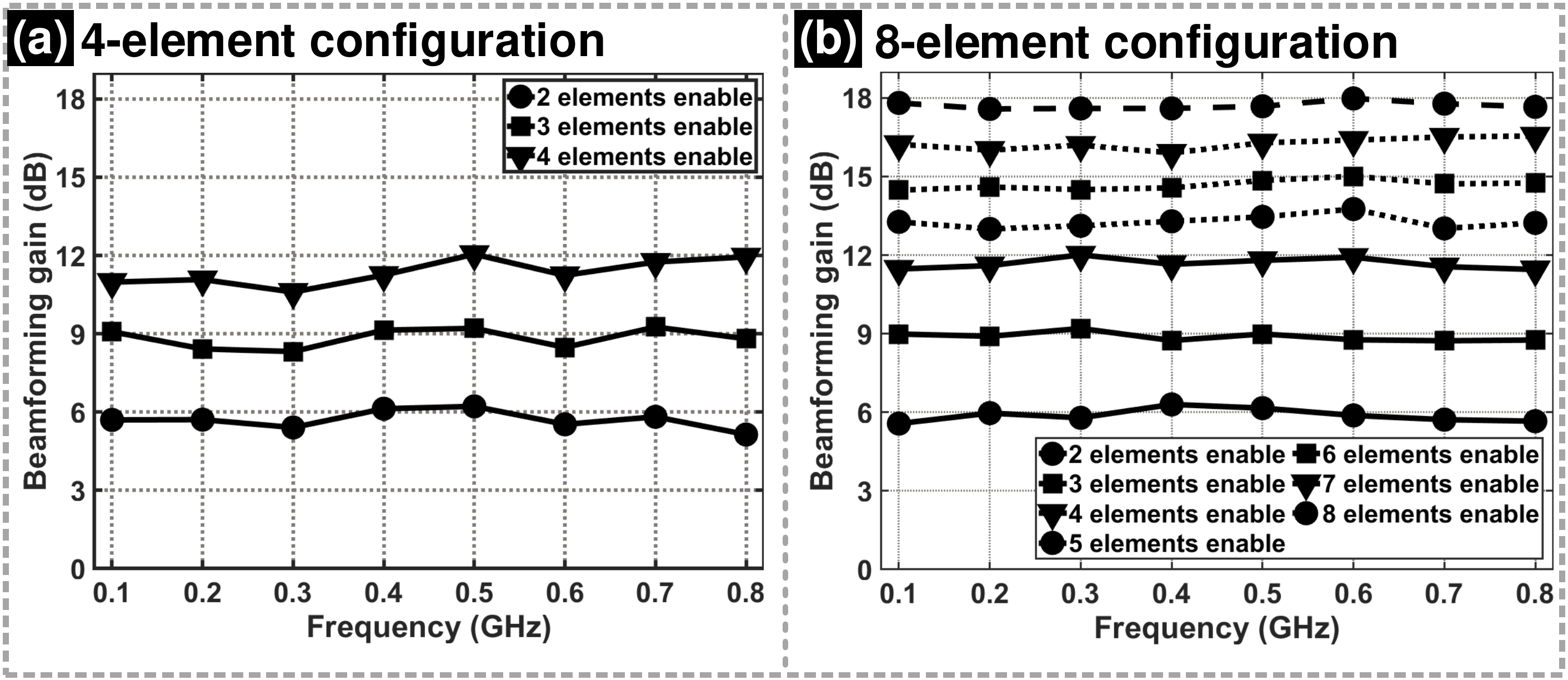}
    \vspace{-4mm}
    \caption{\small Beamforming gain for (a) 4-element; and (b) 8-element configurations.}
    \vspace{-5mm}
    \label{fig:fig17}
\end{figure}

\begin{table*}
\footnotesize {
\caption{\small Comparison with state-of-the-art TTD SSP.}
\vspace{-2mm}
\label{tab:tab2}
\setlength{\tabcolsep}{4pt}
\centering
\begin{tabular}{|c|c|c|c|c|c|c|c|c|c|} 
\hline
\multirow{2}{*}{}                  & \textbf{\cite{chu_2007}}    & \textbf{\cite{cho_2018}}    & \textbf{\cite{mondal_2017}}    &

\textbf{\cite{ghaderi_tcas1_2019}}    & \textbf{\cite{fikes_2021}}       & \textbf{\cite{ghaderi_jssc_2021}}         & \textbf{\cite{spoof_2020}}     & \textbf{\cite{jang_2019}}  & \textbf{This}\\ 
%\cline{2-9}
                                   & \textbf{JSSC’07} & \textbf{RFIC’18} & \textbf{JSSC’17} & 
                                 
                             \textbf{TCAS-I’19} & \textbf{TMTT'21} & \textbf{JSSC’21} & \textbf{SSCL’20} &  \textbf{JSSC’19} & \textbf{work}   \\ 
\hline
%\multirow{2}{*}{\textbf{Arch.}} & \multicolumn{3}{c|}{VCO-CT$\Delta\Sigma$}                              & 
\textbf{TTD Arch.}  &   RF  &   RF  &   BB  &   BB  &   BB  &   BB  &   BB &   Digital &   BB            \\ 
\hline
\multirow{2}{*}{\textbf{Method}}    &   LC      &   LC      &   Gm-C    &   TI          &   TI          &   TI          &   RF          & Digital   &   TI \\
                                    &   delay   &   delay   &   filter  &   sampling    &   sampling    &   sampling    &   resampling  &   delay   &   sampling \\  
\hline
\multirow{2}{*}{\textbf{Domain}}    &   Cont.-      &   Cont.-      &   Cont.-    &   Discrete-          &   Discrete-          &   Discrete-          &   Discrete-          & Digital   &   Discrete- \\
                                    &   Time      &   Time      &   Time    &   Time          &   Time          &   Time          &   Time          &    &   Time \\
\hline
\textbf{\# Elements}    &   4   &   N/A    &   N/A    &   4    &   4   &   4   &   4   &     16  &   4\\
\hline
\textbf{Supply (V)}                   & 1.5 &   2.5 &   1.4 &   1.0 &   N.A.    &   1.0 &   N.A.    &   N.A.    &   1.0/1.3\textsuperscript{1}             \\ 
\hline
\textbf{Power (mW)}                & 555               & 285               & 364            & 47              & 70\textsuperscript{2}               &   40  &   122 & 453             & 29\textsuperscript{3}/176            \\ 
\hline
\textbf{f\textsubscript{CENTER} (MHz)}  &   8   &   11  &   N/A    &   N/A    &   25  &   1.5 &   0.6--4  &   1   &   28\textsuperscript{10}           \\ 
\hline
\textbf{BW (MHz)}                  & 18000  & 2000-20000    & 2000  & 100   &   100 &   500 &   800 & 100   & 800 \\
\hline
\textbf{Delay Range (ns)}                 &    0.3         & 0.508            & 1.7             & 15            & 10                  & 1                & 5             & 7500  &   3.8           \\ 
\hline
\textbf{Delay Resol. (ns)}                & 0.015               & 0.004              & 0.001          & 0.005              & 0.005                  & 0.015                    & 0.035              & 0.25   & 0.005              \\ 
\hline
\textbf{Linearity (dBm)}               & -3.2\textsuperscript{4} (P1-dB)            & N/A            & -3.1 (IIP3)             & N/A         & -28\textsuperscript{5} (EVM)               &  7.9 (IIP3) & -16.5\textsuperscript{6} (IIP3) & -41            & 14\textsuperscript{6} (IIP3)             \\ 
\hline
\textbf{Noise (dB)}                      & 2.9--4.8\textsuperscript{12} (NF)                &  N/A                & 23\textsuperscript{7} (NF)               & 330 $\mu V\textsubscript{rms,out}$    & N/A & 33 (SNDR) & 20\textsuperscript{6} (NF)                & 60\textsuperscript{8} (SNDR) &  31.8\textsuperscript{11}              \\ 
\hline
\textbf{Area (mm\textsuperscript{2})}                & 9.92           & 5.45             & 0.61\textsuperscript{9}            & 0.57\textsuperscript{9}            & 0.5               & 0.82               & 1.2           & 4.42    &   1.98             \\
\hline
\multirow{2}{*}{\textbf{Tech. (nm)}}    &   CMOS    &   BiCMOS  &   CMOS    &   CMOS    &   CMOS    &   CMOS    &   CMOS    &   CMOS    &   CMOS\\      
                                       &    130     &   130     &   130     &   65      &   65      &   65      & 
28      &   40      &   65 \\ 
\hline

\end{tabular}
} \\
\textsuperscript{1}1.3V biases input buffers only; \textsuperscript{2}Baseband delay circuit, mixers, and LO; \textsuperscript{3}Excluding I/O buffers for testing; \textsuperscript{4}Estimated from Fig.15(d) \cite{cho_2018} at 15 GHz; \textsuperscript{5}Test at 270-Mb/s 64-QAM; \textsuperscript{6}Worst case for single-channel measurement; \textsuperscript{7}Worst case with a filter order of 9; \textsuperscript{8}Array SNDR; \textsuperscript{9}Active area only; \textsuperscript{10}Front-end emulated using Xilinx ZCU216; \textsuperscript{11}790MHz input signal with 5dB buffer loss is decoupled; \textsuperscript{12}Meausred from the UWB front-end; * - Estimated\\
\vspace{-8mm}
\end{table*}

 \begin{figure}[t]
\centering
    \includegraphics[width=0.8\columnwidth]{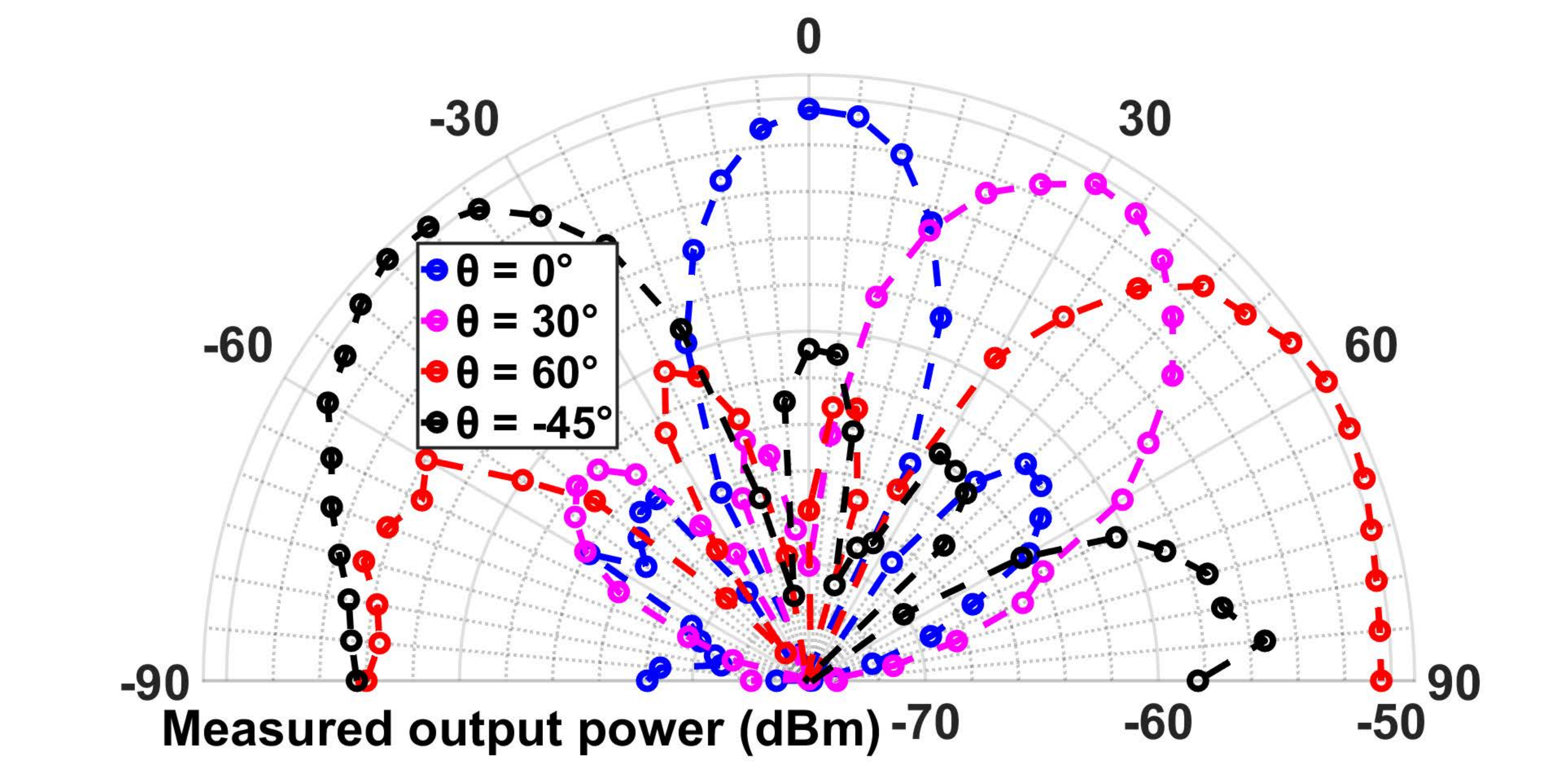}
    \vspace{-3mm}
    \caption{\small AoA measurement in a 4-element configuration.}
    \vspace{-4mm}
    \label{fig:fig18}
\end{figure}

 \begin{figure}[t]
\centering
    \includegraphics[width=1\columnwidth]{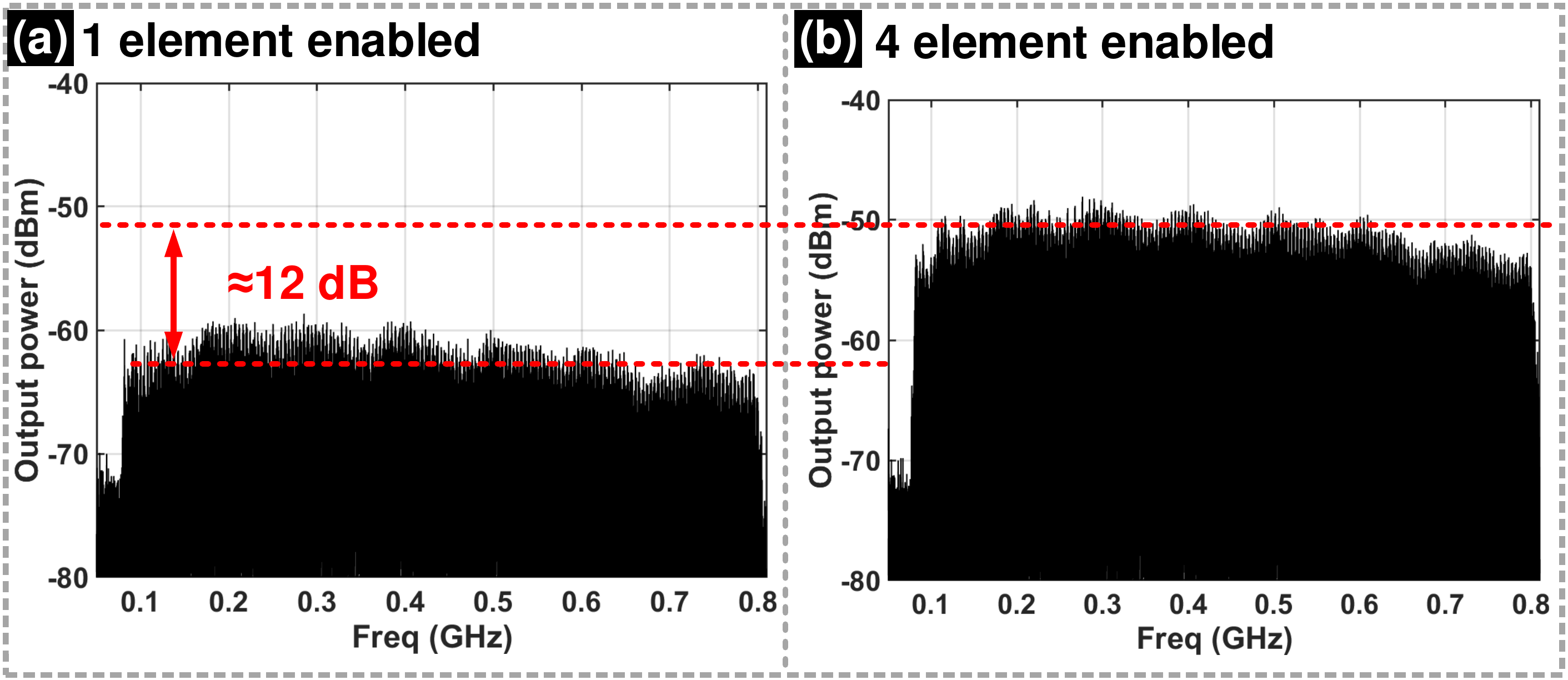}
    \vspace{-4mm}
    \caption{\small Wideband measurement with (a) 1-element; and (b) 4-element enabled.}
    \vspace{-6mm}
    \label{fig:fig19}
\end{figure}

 \begin{figure}[t]
\centering
    \includegraphics[width=1\columnwidth]{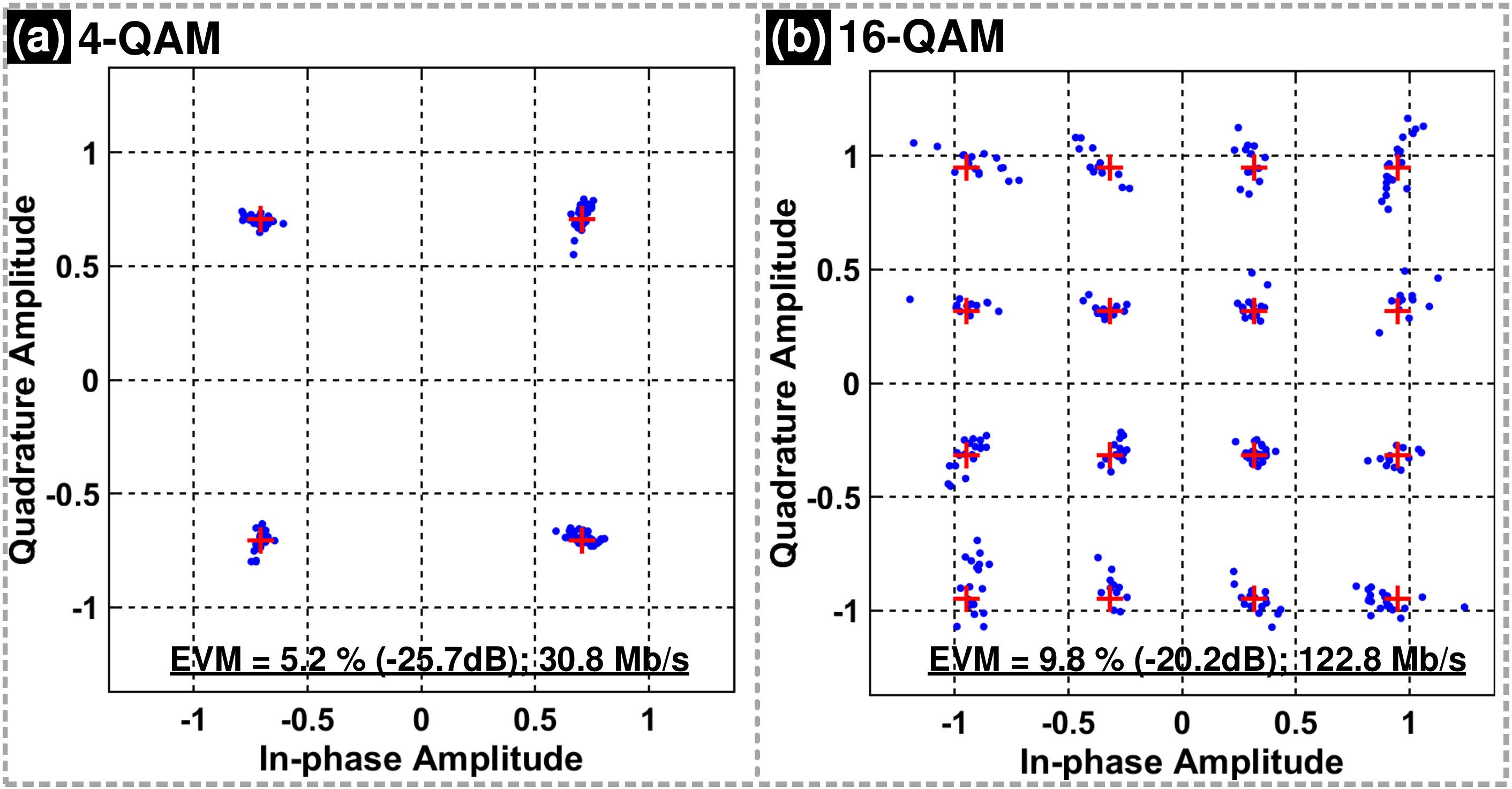}
    \vspace{-4mm}
    \caption{\small Measured EVM for (a) 4-QAM; and (b) 16-QAM.}
    \vspace{-6mm}
    \label{fig:fig20}
\end{figure}

\subsection{Mode 2: Wideband data communications tests}
In the communication mode, the proposed SSP can be configured as illustrated in Fig.~\ref{fig:fig4}(b). Three different signal types are emulated including single tone, wideband chirp, and quadrature amplitude modulation (QAM) in the RFSoC and applied to the TTD SSP DUT. Figure~\ref{fig:fig17}(a) shows the measured single tone response showing a beamforming gain close to 12~dB for a 4-element array with a FoV around 180$^\circ$ across the bandwidth of interest, demonstrating the capability of the delay compensation technique with RAMP based signal combiner across 720~MHz bandwidth. The proposed array can also be configured to 8-element mode, demonstrating a maximum 16~dB beamforming gain, as shown in Fig.~20(b). Though the FoV is decreased to 50$^\circ$ due to the increased number of elements, the narrowed FoV can be remedied by increasing the interleaving level at the expense of additional power consumption and area overhead. For the four cases of angle-of-arrival (AoA), the beamforming beam patterns are presented in the polar format (Fig.~\ref{fig:fig18}). The frequency-independent beamforming conversion gain and the beam-squint free beam patterns show the TTD-based operation of the SSP. Wideband signal testing is also applied to the proposed SSP as shown in Fig.~\ref{fig:fig19}(a)-(b) for 720~MHz BW at 0$^\circ$  AoA with all elements enabled. Again, a uniform beamforming gain of ∼12~dB is observed with all the elements enabled in the 4-element configuration. 
The SSP is also measured for QAM modulated signals with 4-QAM and 16-QAM in the communication mode. The constellation of the desired signal with 4-QAM and 16-QAM input at the DUT output is shown in Fig.~\ref{fig:fig20}(a)(b) respectively. After beamforming, 5.2\%, and 9.8\% EVM is achieved realizing 30.8~Mb/s for 4-QAM and 122.8~Mb/s for 16-QAM respectively. It is also worth mentioning that the source follower buffer contributes additional loss which is not decoupled and thus limits the EVM performance. The single channel third order input intercept (IIP3) is also measured and interpolated for the proposed SSP. Two input tones set to be 766~MHz and 776~MHz with a spacing of 10~MHz results in the third-order intermodulation product located at 786~MHz and 756~MHz respectively demonstrating 14dBm IIP3. The 4-element TTD SSP consumes 29~mW including RAMP, sample phase generation, and the switched capacitor bank. The required interface circuits consume 135~mW, 10~mW, and 2~mW for input buffer, output buffer, and current mirroring respectively. Table~\ref{tab:tab2} summarizes the critical parameters for the proposed TTD SSP and compares with state-of-the-art. Though \cite{spoof_2020} demonstrated similar bandwidth as the proposed work, the use of a RF sampling mixer with digital delay line will limit its application at millimeter-wave. In contrast, the proposed work can connect with different RF front end downconverters relaxing the overall system design complexity for TTD arrays. The modulated signals’ performance in two processing modes proves the applicability of the proposed SSP for high-speed wireless links for both beamtraining and beamforming functionalities.

\section{Conclusions}
This article demonstrates multi-mode SSP with low-latency beamtraining and wideband data communications. Frequency dependent search beams are created to sound all directions simultaneously using true-time-delay arrays to greatly reduce beam training latency. The proposed method is scalable depending on the multi-antenna front-end specifications. Additionally, the proposed architecture supports wideband data communications with large delay-bandwidth product using fast slewing wideband RAMP for efficient signal combining in the baseband switched-capacitor array.   A 3.8ns delay compensation across 800~MHz bandwidth and a 29~mW power consumption is demonstrated with EVM of $<$10\% supporting 16-QAM. 
% or


%\section*{Acknowledgment}


\ifCLASSOPTIONcaptionsoff
  \newpage
\fi



%\bibliographystyle{IEEEtran}

\bibliographystyle{IEEEtran}

\bibliography{IEEEabrv,ref}\textbf{}
\end{document}